\documentclass{aa}

\usepackage[english]{}
\usepackage{natbib}
\usepackage[utf8]{inputenc}
\usepackage[colorlinks,urlcolor=magenta,citecolor=blue,linktocpage,breaklinks]{hyperref}
\usepackage[varg]{txfonts}
\usepackage{adjustbox} 
\usepackage{xspace}
\usepackage[]{footnote}
\usepackage{booktabs}
\usepackage[table,xcdraw]{xcolor}
\usepackage{rotating}
\usepackage{graphicx}
\usepackage{orcidlink}
\usepackage{siunitx}
\begin{document} 

\newcommand{\notee}[1]{\textcolor{purple}{#1}}
\newcommand{\samane}[1]{\textcolor{blue}{#1}}

   \title{Revisiting the structure of galactic disks with deep imaging\thanks{We dedicate this work to the memory of Peter Erwin, a pioneer in the classification of disk galaxy types, who unfortunately left us too soon.}}

   \author{Samane Raji\orcidlink{0000-0001-9000-5507}\inst{1\textbf{, 2}}, Ignacio Trujillo \orcidlink{0000-0001-8647-2874}\inst{4,5},  Fernando Buitrago\orcidlink{0000-0002-2861-9812}\inst{1\textbf{, 2},3}, Giulia Golini\orcidlink{0009-0001-2377-272X}\inst{4,5} and Ignacio Ruiz Cejudo\orcidlink{0009-0003-6502-7714}\inst{4,5} }

   \institute{Departamento de Física Teórica, Atómica y Óptica, Universidad de Valladolid, 47011, Valladolid, Spain
   \and Laboratory for Disruptive Interdisciplinary Science (LaDIS), Universidad de Valladolid, 47011 Valladolid, Spain
   \and Instituto de Astrofísica e Ciéncia do Espaço, Universidad de Lisboa, OAL, Tapada de Ajuda, PT1349-018 Lisbon, Portugal
   \and Instituto de Astrofísica de Canarias,c/ Vía Láctea s/n, E38205 - La Laguna, Tenerife, Spain 
   \and Departamento de Astrofísica, Universidad de La Laguna, E-38205 - La Laguna, Tenerife, Spain
  }

  \date{Received July 18, 2025; accepted Octobre 24, 2025}

 
  \abstract
  {Thanks to new advances in astronomical imaging, we can now routinely explore disk galaxy profiles about two magnitudes deeper than the data available 20 years ago. In this regard, it is an opportune time to reevaluate the past classifications of different surface brightness Types. In this paper, we explore the stellar mass profiles of a sample of disk galaxies with similar stellar masses ($\sim$10$^{10}$ M$_{\odot}$) using IAC Stripe82 Legacy Project data. We find that Type I, II-CT, and III-d galaxies exhibit edges at surface brightnesses $\mu_g$(R$_{edge}$)$\sim$26.5 mag/arcsec$^2$ and surface mass densities $\Sigma$(R$_{edge}$)$\sim$0.5-1 M$_{\odot}$/pc$^2$. These surface brightnesses were outside the range of typical SDSS images and, therefore, unstudied. The present data suggest (although the statistics are inconclusive) that the main difference between the previous profile Types is the presence of either a more or a less intense bulge. Half of the Type II-CT galaxies are Sc, whereas half of the Type III-d galaxies are Sb. Finally, Type III-d galaxies have low mass surface density disks ($\Sigma$($\Delta$ $R_{\mathrm{exp}}$)$\sim$0.5 M$_{\odot}$/pc$^2$) and are bluer $(g-r)_0$=0.32$\pm$0.02 than the other Types ($(g-r)_0$=0.42$\pm$0.02 for Type I and II-CT), suggesting that their disks probably formed later in time.}

\keywords{galaxies: photometry – galaxies: structure – galaxies: fundamental parameters – galaxies: evolution – galaxies: formation}

\titlerunning{Revisiting the structure of galactic disks}

\authorrunning{Raji et al.}

   \maketitle

\section{Introduction}

About 20 years ago, the ability to characterize the surface brightness profiles of around a hundred galaxies using wide field optical surveys such as the Sloan Digital Sky Survey  \citep[SDSS; ][]{2000AJ....120.1579Y} allowed disk galaxies to be grouped into three basic Types \citep{Erwin2005,Pohlen_2006}. These basic Types are: pure exponential profiles with no breaks (Type I), Type II with a "downbending break" (revising and extending a classification introduced by \citet{1970ApJ...160..811F} to include the so-called truncation (edge) of the disk discovered by  \citealt{1979A&AS...38...15V}), and what was then a completely new class (Type III), also described by a broken exponential but with an upbending profile \citep{Erwin2005}.

Over time, the physical interpretation of each profile type (except Type I) has become more settled. From the beginning, Type III profiles have been divided into two subgroups, depending on whether the excess light in the outer part of the galaxy can be attributed to a stellar halo (i.e. Type III-s) or to a stellar disk in the process of formation (i.e. Type III-d). Star formation in the outer regions should produce an excess of light over the internal exponential decay \citep{Erwin2005}. This scenario was further supported by the finding that Type III galaxies have irregularities in the outer part, suggesting a recent merger event activating star formation in the outermost part of the galaxy \citep{pohlen2006}.  

In the case of Type II, the most widely accepted interpretation was given by \citet{2008ApJ...675L..65R}. They found that an idealized galaxy formed by the dissipative collapse of gas embedded in a spherical dark matter halo would exhibit a broken stellar surface density profile (Type II). In this scenario, the radial position of the edge\footnote{\citet{2008ApJ...675L..65R} used the term "break" to describe what we now call an "edge." Currently, "break" refers to a change in surface brightness profiles, while "edge" refers to a drop in stellar surface mass density. In some galaxies, both features coincide, but in many, breaks are only seen in surface brightness profiles at radial distances closer to the center of the galaxy than the edges \citep{martin-navarro2012}.} corresponds to the location of a rapid drop in the star formation rate associated with a sudden decrease in the surface density of the cooled gas. The origin of stars beyond the edge results from the outward scattering (i.e., migration) of stars from the inner disk. As a result of this evolutionary process, the radial mean stellar age profile of the galaxy is U-shaped, with the younger ages located at the edge of the star-forming disk \citep{Debattista2006,2008ApJ...675L..65R,martinez2009,sanchez2009}.

The theoretical U-shape prediction of \citet{2008ApJ...675L..65R} was almost immediately supported by observations. Using the optical color profile of galaxies as a proxy for their age profiles, analysis of the color profiles showed that their shape was indeed U-shaped. This was found both in disk galaxies at intermediate redshifts and in the present-day universe \citep{Azzollini2008,Bakos2008,bakos2012,martin-navarro2012,Zheng2015,Watkins_2016,Ruiz-Lara2016}.  In support of the scenario presented by \citet{2008ApJ...675L..65R}, the position of the minimum of the U-shape was in qualitative agreement with the radial position of the break in the surface brightness profiles. Given the modest depth ($\mu_{g,lim}$$\sim$26.5 mag/arcsec$^2$) of these initial observations, not much light was visible beyond the break of the galaxy, strengthening (at that time) the interpretation that the minimum of the U-shape corresponds to the visual end of the star-forming disk.

The origin of the purely exponential (i.e. Type I) profiles remains a mystery since then. Considering that star formation seems to be efficient only above a certain gas density threshold and/or that galaxies undergo accretion processes with other smaller galaxies, there is no obvious physical reason why the surface brightness profile of disk galaxies should continue to decay in this way to fainter and fainter surface brightnesses\footnote{A similar argument applies to Type III-d galaxies. If their outer excess of light corresponds to a young star-forming disk, we would expect that sufficiently deep imaging would reveal the edge of the galaxy.}. This, together with the fact that this type of profile is in the minority, with only 10\% of cases, suggests that it is only the shallow surface brightness limit of surveys such as the SDSS that has prevented finding the galaxy edge in this type of case.

In recent years, there has been a revolution in the depth at which galaxies can be imaged. Today, multiple works are able to obtain images 2 to 3 mag deeper (i.e. $\mu_{g,lim}$$\sim$29-30 mag/arcsec$^2$) than those obtained by the SDSS survey \citep[see e.g.][]{2012ApJS..200....4F,2013ApJ...762...82M,2015MNRAS.446..120D,2015A&A...581A..10C,2016MNRAS.456.1359F,2016ApJ...830...62M,2018MNRAS.480..521H,2018ApJ...857..144H,2019MNRAS.490.1539R,2024A&A...684A..99G}. These give us the opportunity to explore the structure of galaxies in regions that were hidden in the past in more detail. Moreover, one of the major differences from the initial work on galaxy profile classification is the increasing use of stellar mass surface density profiles, in addition to or instead of surface brightness profiles. The stellar mass surface density profiles are less susceptible to internal substructures, such as star-forming regions and spiral arms, which are better traced by surface brightness profiles (mainly in the bluer bands). 
As a result, galaxies with clear breaks in their surface brightness profiles (i.e., Type II) show a very mild (if any) break when explored using stellar surface mass density \citep[see e.g.][]{Bakos2008,2012MNRAS.427.1102M}. 

Finally, the increasing depth has opened the possibility of exploring the edges of galaxies, which we understand here as the further radial location where there is a significant drop in current or past in situ star formation \citep{Trujillo2020}. This physical phenomenon leaves an imprint in the form of a drop in the stellar mass density profile. These edges have been successfully identified in both nearby \citep[see e.g.][]{Chamba_2022,2025A&A...700A..91G}  and intermediate-z galaxies \citep{Buitrago2023}. With all this new information, now is an opportune time to return to the classification of disk galaxy profile Types from 20 years ago and determine if our interpretation of the origin of the different classes needs updating.

This work revisits the classification of disk galaxy profiles in light of what we have learned over the past two decades through deep imaging and stellar mass surface density profiles. Specifically, we are interested in analyzing whether the galaxies with different surface brightness profile Types exhibit edges in the mass density profiles and if a unifying framework can be established when exploring the different galaxy types from this new perspective.

To address the above questions, we use data from the IAC Stripe82 Legacy Project \citep{2016MNRAS.456.1359F,2018RNAAS...2..144R} to analyze the structure of a sample of disk galaxies with stellar masses around 10$^{10}$ M$_{\odot}$ \citep{Trujillo2020,Chamba_2022}. We will present evidence that current deep imaging supports the idea that, although they have different surface brightness profiles, Type I and Type II-CT disk galaxies look much the same when explored using their stellar mass surface density profile. Both types have an exponential stellar mass profile describing their disk, followed by a downward-bending profile at their edge. For Type III-d galaxies, there is no obvious edge in their mass profile when using elliptically averaged profiles. However, we will demonstrate that this is related to the significant asymmetry of their star-forming disks. When using wedge profiles, Type III-d galaxies also present an edge.

The paper is structured as follows. Section \ref{data} describes the sample used and how the different galaxy profiles were classified. In Section \ref{analysis}, we explain how the average stellar mass and color profiles were created and which properties they have. Finally, Section \ref{summary} analyzes and summarizes the main findings of this work. A standard $\Lambda$CDM cosmology with $\Omega_m = 0.3$, $\Omega_\Lambda = 0.7$, and $H_0 = 70 \ \mathrm{km\ s^{-1}\ Mpc^{-1}}$ was assumed in this paper. Magnitudes are provided in the AB system \citep{oke1983}.

\section{Data and sample selection}
\label{data}

\subsection{Deep Stripe82 imaging}

The galaxies analyzed in this paper were obtained from the parent sample described by \citet{Trujillo2020}. This sample consists of 1005 galaxies (279 ellipticals, 464 spirals, and 262 dwarfs) with stellar masses between 10$^{7}$ to 10$^{12}$ M$_{\odot}$ and redshift 0.01$<$z$<$0.1. For the purpose of this work, we concentrate only on the spiral galaxies with morphological types between 1 to 7 (i.e., from Sa to Sd). This left a total of 227 objects. The profiles analyzed in this work were obtained using images from the IAC Stripe82 Legacy Project \citep{2016MNRAS.456.1359F,2018RNAAS...2..144R}. This project is based on a co-added version of the SDSS  ‘Stripe 82’ \citep{2008AJ....135.1057J,2009ApJS..182..543A} that has been optimized for low surface brightness astronomy. The limiting depths in surface brightness of these images are $\mu_g$=29.1 mag/arcsec$^2$ and $\mu_r$=28.5 mag/arcsec$^2$,  both measured as a 3$\sigma$ fluctuation with respect to the background of the image in areas equivalent to 10\arcsec$\times$10\arcsec\ boxes. The g and r bands surface brightness and the stellar mass profiles of the galaxies were obtained by \citet{Chamba_2022}. While \citet{Chamba_2022} provides a detailed description of the methodology used to obtain these profiles, here we offer a concise summary. Before extracting the profiles, the neighboring sources are carefully masked using Max Tree Objects \citep[MTO; ][]{Teeninga2016,2021A&A...645A.107H}, and the scattered light produced by the brightest source in the image is removed following the procedure described in \citet{infante2020}. The background of the image is calculated using the masked image and is subtracted from the data. The center, axis ratio, and position angle of the ellipses used to extract the galaxy profiles are fixed and correspond to those of isophote 26 mag arcsec$^{-2}$ in the g-band. The uncertainties in the surface brightness profiles are estimated at each radial position as the sum of the uncertainty in the background estimation and the Poissonian error associated with the flux of the isophote at that distance. The surface brightness profiles are corrected for inclination effects, redshift dimming, and Galactic extinction. \citet{Chamba_2022} compute the stellar mass density profiles following the prescription given in \citet{2015MNRAS.452.3209R} and apply the set of equations provided in \citet{Bakos2008}. The basis for the stellar mass density profile is the deepest band; in their case, this is the g-band.

To have a representative stellar mass sample, we select only those disk galaxies within a narrow stellar mass range around 10$^{10}$ M$_{\odot}$. In particular, we take galaxies with 10$^{9.8}$ to 10$^{10.2}$ M$_{\odot}$. This range corresponds to a factor of around two in mass, which is similar to the uncertainty in estimating stellar masses based on the stellar population properties \citep[see e.g.][]{2014ApJS..210....3M}. Our mass range results in a sample of 44 galaxies (see Table \ref{galaxyproperty}).  The selection of the stellar mass around 10$^{10}$ M$_{\odot}$ is done for several reasons. First, this stellar mass roughly corresponds to the peak of the stellar mass distribution of the disk galaxies \citep[see e.g.][]{2014MNRAS.444.1647K}. Therefore, by selecting this mass range, we explore the disk galaxies that are more common. Second, the construction of the original galaxy sample is biased in that lower-mass galaxies are closer in distance than higher-mass galaxies. For example, the average redshift of 10$^{10}$ M$_{\odot}$ galaxies is about 0.02, while that of 10$^{10.5}$ M$_{\odot}$ galaxies is 0.04. Therefore, the spatial resolution of the surface brightness and stellar mass profiles is better by a factor of 2 for lower stellar mass galaxies. Third, the range of morphological T-types selected by disk galaxies in the previous mass interval is between 1 and 6. This fits well with the sample studied by \citet{pohlen2006}, which covers basically the same morphological T-types. Finally, the lower the stellar mass of the galaxies, the less likely the merger activity is to alter the structural properties of the galaxies, allowing a clearer analysis of the disk structure of these objects.

\subsection{Galaxy profile classification}
\label{classification}

We have classified the surface brightness profiles of the 44 selected disk galaxies as Type I, II, or III, following the same scheme\footnote{\citet{pohlen2006} sub-classified Types II and III into several classes. We have followed that classification scheme here as well. Type II-CT refers to "Classical Truncations," i.e., those compatible with the location of truncations found in edge-on galaxies \citep{1979A&AS...38...15V}. Type II-AB refers to a break that can be associated with galaxies that have an Apparent or Asymmetric Break generated by the galaxy's different extensions in different directions. Type OLR is used when the galaxy has a bar and the break is connected to an Outer Lindblad Resonance (i.e., located at twice the bar radius). When a bar is present, Type II.i and Type II.o refer to breaks inside and outside the bar extension, respectively. For Type III, the addition of "d" or "s" refers to the origin of the excess light beyond the edge, which is produced by either a disk-like or stellar-halo-like structure. Finally, the "Others" category contains profiles that cannot be associated with the previous categories due to their mixed shapes. In this work, for simplicity, we also included in the category "Others" the Types II.o-OLR and II.i.} as \citet{pohlen2006}. For this reason, to perform the classification of each surface brightness profile, we looked only at the shape of the profile down to a surface brightness of 26.5 mag/arcsec$^2$ (r-band), which is roughly the typical SDSS depth of the surface brightness profiles analyzed in \citet{pohlen2006}. In other words, to be consistent with earlier work, our classification of the surface brightness profiles by eye neglects the shape of the profiles beyond the SDSS surface brightness limit. The description of each galaxy, together with an explanation of the reasons for the choice of profile Type, is given in the Section \ref{galaxydescription}.  We show all the surface brightness profiles in g and r bands, together with the radial color and stellar mass profiles, in the Appendix \ref{profiles}.

\subsection{Sample description: morphological and profile Types} 
\label{galaxydescription}

We give a brief description of the properties of the galaxies analyzed in this paper. In addition, we give the reasons for the classification of the profile Type.

\paragraph{UGC 139}, also known as CGCG 382-035, is a Scd galaxy according to \citet{nair}. The galaxy has a prominent star-forming disk with a very blue color ($g-r$)$_0$$\sim$0.2 mag. The position of the break of this galaxy coincides with the position of the edge. Based on the shape of the profile, we classify the galaxy as Type II-CT. A galaxy with a similar surface brightness profile in \citet{pohlen2006} is UGC 09837. 

\paragraph{MCG+00-01-057} has an Sbc morphology according to \citet{nair}.  The profile of the galaxy shows a break at about R/$R_{edge}$=0.4. This break seems to be associated with an inner ring-like structure and coincides with the minimum color profile. We classify the profile as Type II.i. A galaxy with a similar surface brightness profile in \citet{pohlen2006} is UGC 06518.

 \paragraph{MCG+00-02-081} is a Sc galaxy \citep{nair}. We classify the galaxy profile as II-CT due to the presence of a downbending break at R/$R_{edge}$=0.5. This break coincides with the minimum of the ($g-r$)$_0$ color profile.   This galaxy has similar morphological and profile properties to those of NGC 2684 shown in \citet{pohlen2006}.
 
\paragraph{MCG+00-02-126} has a Sc morphology \citep{nair}. The object has an interaction with a nearby neighbor. This interaction washes out the possibility of identifying a clear edge.  Given the asymmetry of the outer part and the presence of a break in the surface brightness profiles at about $R/R_{edge}$ = 0.4, we have classified the profile as Type II-AB. The minimum of the color profile lies at the break radius. This galaxy has similar properties to NGC 6155 shown in \citet{pohlen2006}. 

\paragraph{UGC466} has a Sc morphology \citep{nair}. The galaxy is not symmetric; the northern side is more extended than the southern side. Down to a surface brightness of $\sim$26.5 mag/arcsec$^2$ (i.e., approximately the SDSS surface brightness limit), the galaxy shows a gently decreasing exponential behavior, and we therefore classify its profile as Type I. A galaxy with similar profile characteristics in \citet{pohlen2006} is IC 1125.

\paragraph{UGC737} is an Sb galaxy \citep{nair}. We classify the surface brightness profile of this galaxy as Type III-d. This galaxy has a distinctly blue ($g-r$)$_0$=0.35 disk where the reddening begins beyond the $R_{edge}$. A galaxy with a similar surface brightness profile is UGC 09741 in \citet{pohlen2006}. 

\paragraph{MGC+00-04-086} is a red compact disk galaxy with a stellar halo-like structure surrounding it, probably indicating an ongoing minor merger. It has a morphology of Scd \citep{nair}. The profile shows an up-bending in the outer region. We classify the profile as Type III-s. A similar surface brightness profile can be found in the study of \cite{pohlen2006} for the galaxy UGC 10721. 

\paragraph{MGC+00-04-126} has a morphology of Sc \citep{nair}. Its profile shows a break at about half the location of $R_{edge}$. The minimum of the color profile is at the same location as the break. Beyond the break the color starts to get redder. We classify this profile as Type II-CT.  A galaxy with a similar profile in \cite{pohlen2006} is NGC 3589.

\paragraph{MCG+00-04-132} is an Sbc galaxy according to \citet{nair}. The galaxy has an up-bending profile. The excess of light in the outer part is associated with the presence of recent star formation in the outer part. Around $R_{edge}$, the profile of the galaxy becomes redder, indicating a possible end of in situ star formation of the object. We classify the profile as Type III-d. A galaxy in \cite{pohlen2006} with a similar surface brightness profile is UGC 10721.

\paragraph{LEDA1137500} has an Sa morphology \citep{nair}. The excess of light in the outer part of the galaxy is probably related to the presence of a stellar halo, although we cannot exclude the possibility that it could be explained by the effect of scattered light from the (inclined) galaxy itself. We classify the profile as Type III-s. A galaxy with a similar profile in \cite{pohlen2006} is NGC 1084.

\paragraph{IC138} is an Sbc galaxy \citep{nair}. The surface brightness profile has a break at $R/R_{edge}$=0.75. Beyond this location, there is an increasing reddening of the color profile. We classify this profile as Type II-CT. A galaxy in \cite{pohlen2006} with a similar surface brightness profile is NGC 7437.

\paragraph{MGC+00-05-013} is a Sab galaxy according to 
 \citet{nair}. The galaxy shows the presence of a bar and an elongated ring-like structure surrounding it. This causes an excess of light in the profile of the galaxy at $R/R_{edge}$=0.35. This ring is bluer than the rest of the galaxy. We classify the profile as II.o-OLR. A galaxy with a similar surface brightness profile in \cite{pohlen2006} is NGC 5850.

\paragraph{MGC+00-05-045} has an Sc morphology according to \citet{nair}. There are two possible locations for the break in the galaxy. One at $R/R_{edge}$=0.4 and another at 0.8. This radial range also coincides with the bluer part of the galaxy. We classify the profile as Type II-CT. A galaxy with a similar surface brightness profile in \cite{pohlen2006} is NGC 3589. 

\paragraph{UGC1603} has been classified as Sc by \citet{nair}.
The galaxy has a surface brightness profile with an exponential decrease compatible with a Type I. The color of the galaxy is red in the center and blue from $R/R_{edge}$=0.3 to the end of the observable disk. A galaxy with a similar surface brightness profile in \cite{pohlen2006} is NGC 5668. 

\paragraph{2MASXJ02135431+0026533} has a Sa morphology \citep{nair}. The galaxy has a ring-like structure. The profile of the galaxy shows a notable color change from red to blue at about $R/R_{edge}$=0.5. This coincides with the end of the ring-like structure and reinforces the idea that the outer part of the galaxy is probably affected by the resonance produced by the inner ring, causing an increase in star formation outside. We classify its profile as Type II.o-OLR.  A galaxy in \cite{pohlen2006} with a similar surface brightness profile is NGC 1068.

\paragraph{MCG+00-07-031} is a low inclination galaxy with Sb morphology \citep{nair} and profile Type III-d. We make this profile classification based on the outer region of the galaxy, which has a spiral structure with blue colors. A galaxy in \cite{pohlen2006} with a similar surface brightness profile is NGC 2967. 

\paragraph{UGC2019} has an Sc morphology according to \cite{nair} and we classify its profile as Type II-CT. The break of the galaxy is at $R/R_{edge}$=0.5. A galaxy in \cite{pohlen2006} with a similar profile is NGC3586. The galaxy has a non-symmetric outer region that obscures a clear identification of the edge. 

\paragraph{LEDA175542} is an Sb galaxy \citep{nair} with a ring-like structure surrounding the bulge region. For this reason, we classify its profile as Type II-OLR. The color profile has its minimum at $R/R_{edge}$=0.3. A galaxy with a similar surface brightness profile in \cite{pohlen2006} is NGC 5376.

\paragraph{LEDA11066} is an Sb galaxy \citep{nair} that has a ring-like structure surrounding its bulge. We classify its surface brightness profile as Type II-OLR. A galaxy with a similar surface brightness profile in \cite{pohlen2006} is NGC5376. The outermost region of this galaxy is not very symmetric, perhaps indicating recent merging activity. The edge of the galaxy is probably washed out by this interaction.
\paragraph{KUV02558+0026} is a galaxy with a non-symmetric outer part, classified as Sc by \citet{nair}. We classify the surface brightness profile as Type III-s, since the excess of light in the outer part may be related to a recent interaction. A galaxy in \cite{pohlen2006} with a similar surface brightness profile is NGC 5624. 

\paragraph{LEDA1133341} is an asymmetric galaxy with a morphological type Sc according to \citet{nair}. Due to the break in the surface brightness profile and the asymmetry of the galaxy, we classify the surface brightness profile of this galaxy as Type II-AB. The break of the galaxy is located at about $R/R_{edge}$=0.5 and coincides with the end of the bluer part of the object. The possible recent merger activity of this galaxy has washed out the possibility of detecting a clear edge. A galaxy with a similar surface brightness profile in \citet{pohlen2006} is NGC 3246. 

\paragraph{LEDA145248}: is a galaxy with an Sbc morphology \citep{nair}. Down to a surface brightness of about 26.5 mag/arcsec$^2$ (i.e., approximately the SDSS surface brightness limit), the galaxy has an exponentially decreasing profile. Therefore, we classify the galaxy profile as Type I. A galaxy with a similar surface brightness profile in \citet{pohlen2006} is NGC 2776.

\paragraph{LEDA1145667} has a Sab morphology according to \citet{nair}. The profile shows a subtle break at about $R/R_{edge}$=0.8. However, up to the SDSS surface brightness limit, the surface brightness profile would have been classified as Type I. We therefore classified it as such. A galaxy in \citet{pohlen2006} with a similar profile is NGC 2776.

\paragraph{Z390-4} is a galaxy whose non-symmetric outskirts suggest that it is undergoing a minor merger. The morphology of the galaxy is Sab according to \citet{nair}. The surface brightness profile shows a break at $R/R_{edge}$=0.4, which is also associated with a transition in the color profile. The profile has an excess of light in the outer part, which is most likely related to the merging activity, and therefore, we classify the galaxy profile as Type III-s. A galaxy with a similar surface brightness profile in \citet{pohlen2006} is NGC 0853. 

\paragraph{Z390-17} is a low inclination disk galaxy of morphological type Sab according to \citet{nair}. The profile of the galaxy shows a break at about $R/R_{edge}$=0.8. At this radius, the color profile becomes redder. We classify the profile as Type II-CT. A galaxy in \citet{pohlen2006} with a similar surface brightness profile is NGC 5660. 

\paragraph{KUG0315+007} is an Scd galaxy according to \citet{nair}. The galaxy has an asymmetric edge. The profile shows a break at about $R/R_{edge}$=0.5, which coincides with a minimum in the color profile. Due to the asymmetry of the galaxy, we classify the profile as Type II-AB. The edge of the galaxy is not very well defined in the profile due to the irregular outer region of the galaxy. An object in \citet{pohlen2006} with a similar surface brightness profile is NGC 6155.

\paragraph{Z390-79} is a low inclined galaxy of the morphological type Scd \citep{nair}. The surface brightness profile shows a break at about $R/R_{edge}$=0.6. The color profile remains blue up to the edge, where it starts to become redder. We classify this profile as Type II-CT. A galaxy with a similar surface brightness profile in \citet{pohlen2006} is NGC 4545.

\paragraph{LEDA176187} is a barred galaxy classified as Sb in \cite{nair}. The outer region of the galaxy suggests a merger in progress. In addition, the profile of the galaxy has an up-bending profile, and we classify the object as Type III (most likely Type III-s). The profile also shows a bump at about $R/R_{edge}$=0.4, which is likely related to the bar structure. A galaxy in \citet{pohlen2006} with a similar profile is NGC 1299.

\paragraph{LEDA176224} is of morphological type Sb \citep{nair}. The galaxy profile shows a break at about $R/R_{edge}$=0.5. The break occurs at a relatively bright surface brightness, and the minimum in the color profile is at $R/R_{edge}$=0.3. For this reason, we classify the profile as an inner break, i.e., type II.i. The edge of the galaxy has been washed out due to the presence of strong Galactic cirrus in the field of this object. A galaxy in the work of \citet{pohlen2006} with a similar surface brightness profile is UGC 06518.

\paragraph{Z391-7} it has an Sc morphology following \citet{nair}. The surface brightness profile shows a break at about $R/R_{edge}$=0.6, where the minimum of the color profile is also located. We classify the profile of the galaxy as Type II-CT. A galaxy with a similar surface brightness profile in \citet{pohlen2006} is NGC 3589. 

\paragraph{LEDA1129114} is an Sb galaxy according to \citet{nair}. The presence of a near compact object complicates a clean analysis of the outer part. The profile has two bumps, one associated with an inner compact disk-like structure and then an arm-like structure at about $R/R_{edge}$=0.8. The formation of these structures could be related to a dynamical resonance, and therefore, we classify the profile as Type II.o-OLR. In \citet{pohlen2006}, a galaxy with a similar profile is NGC 5430.

\paragraph{Z374-27} is a Sc galaxy according to \citet{nair}. The roundish outer region of the galaxy suggests an ongoing merger, although there are some compact objects surrounding the galaxy that make this analysis difficult. Due to the up-bending of the profile, we classify this galaxy as Type III-s. A galaxy in \citet{pohlen2006} with a similar surface brightness is NGC 5624.

\paragraph{Z375-11} is an Scd galaxy \citep{nair}. The profile is compatible with an exponentially decreasing surface brightness down to the SDSS limit (i.e,. about 26.5 mag/arcsec$^2$). For this reason, this object is classified as Type I. We note, however, that there is some evidence for an excess of light beyond $R_{edge}$. The presence of a nearby star (although masked) may be responsible for such an excess, while at the same time washing out a clear detection of the edge of the galaxy in the profile. A galaxy with a similar profile Type in \citet{pohlen2006} is NGC 2776.

\paragraph{Z375-24} has a morphological classification of Sb \citep{nair}. The profile is well described with an exponential decrease down to $R_{edge}$, and for this reason, we classify the galaxy as Type I. A galaxy with a similar profile type in \citet{pohlen2006} is NGC 2776.

\paragraph{MCG+00-54-023} has a morphological type of Sc according to \citet{nair}. The galaxy shows a clear break at about $R/R_{edge}$=0.6, which coincides with the minimum in the color profile. For this reason, we classify the profile of the galaxy as Type II-CT. A galaxy in \citet{pohlen2006} with a similar surface brightness is NGC 3589. 

\paragraph{Z375-48} is a low inclined galaxy of morphological type Sc \citep{nair}. The break of the galaxy is located at $R/R_{edge}$=0.85 and coincides with the color minimum. We classify the profile of the galaxy as Type II-CT. A galaxy with a similar surface brightness profile is NGC 5660 \citep{pohlen2006}.

\paragraph{LEDA1128634} has been morphologically classified as Sc by \citet{nair}. The profile shows a break at about $R/R_{edge}$=0.75 and the characteristic U-shape for the color profile. We classify its profile as Type II-CT. The galaxy NGC 12709 in \citet{pohlen2006} has a profile with similar properties.

\paragraph{SDSSJ221223.75-004745.3} is a Sc galaxy \citet{nair}. It has a break in the profile at about $R/R_{edge}$=0.75. This break coincides with the minimum color of the $g-r$ profile. We classify the profile of the galaxy as Type II-CT. A galaxy with a similar surface brightness profile in \citet{pohlen2006} is UGC 09837.

\paragraph{UGC12446} is a low-inclination Sc galaxy according to \citet{nair}. Although the innermost disk of the galaxy looks quite symmetrical, the outermost region is not. This may indicate an ongoing minor merger. This is also supported by the structure of the surface brightness profile, which shows a break at $R/R_{edge}$=0.75 and an up-bending at $R_{edge}$. For this reason, we classify the galaxy as Type II-CT+III. A galaxy with a similar surface brightness profile in \citet{pohlen2006} is NGC 5667. 

\paragraph{2MASXJ23154453+0027383} has an Scd morphology according to \citet{nair}. The outer region shows an excess of light associated with an active star-forming disk. For this reason, we classify the profile of the galaxy as Type III-d. A galaxy in \citet{pohlen2006} with a similar surface brightness profile is NGC 2967.

\paragraph{UGC12635} is a low inclined galaxy with an Sb morphology \citep{nair}. The surface brightness profile has a break at about $R/R_{edge}$=0.75. For this reason, we classify the profile of the galaxy as Type II-CT. IC 1158 is a galaxy with a similar surface brightness profile in \citet{pohlen2006}. 

\paragraph{LEDA135904} is a low inclined galaxy with an Scd morphology \citep{nair}. The surface brightness profile has a break at about $R/R_{edge}$=0.75. We classify the profile of the galaxy as Type II-CT. A galaxy with a similar surface brightness profile in \citet{pohlen2006} is IC 1158.

\paragraph{MCG+00-60-032} is of morphological type Scd according to \citet{nair}. The surface brightness profile has a break at about $R/R_{edge}$=0.65. We classify the profile of the galaxy as Type II-CT. In this case, however, the minimum of the color profile does not coincide with the location of the break. The galaxy UGC 08658 in \citet{pohlen2006} has a similar surface brightness.

\paragraph{Z382-5} is a non-symmetric disk with a morphological class of Sbc \citep{nair}. The surface brightness profile has a break at about $R/R_{edge}$=0.65. We classify the profile of the galaxy as Type II-AB. A galaxy with a similar surface brightness profile in \citet{pohlen2006} is NGC 6155.

We find the following number of Types: 6 Type I (13$\pm$5\%), 28 Type II (64$\pm$12\%), and 10 Type III (23$\pm$7\%). These numbers are within the errors in nice agreement with those given in \citet{pohlen2006}. The Type II galaxies are sub-classified (again following the scheme by \citealt{pohlen2006}) in 17 Type II-CT, 2 Type II.i, 4 Type AB, and 5 Type II.o-OLR. In the case of Type III, we have the following subclassification: 4 Type III-d and 6 Type III-s.  

\section{Analysis}
\label{analysis}

After analyzing and classifying the surface brightness profiles of the galaxies into different Types, we use the surface brightness, stellar mass density, and $g-r$ color profiles to see what deeper data adds to the classification scheme proposed 20 years ago.

\begin{figure}
    \centering
    \includegraphics[width=1\linewidth, height=1.5\linewidth]{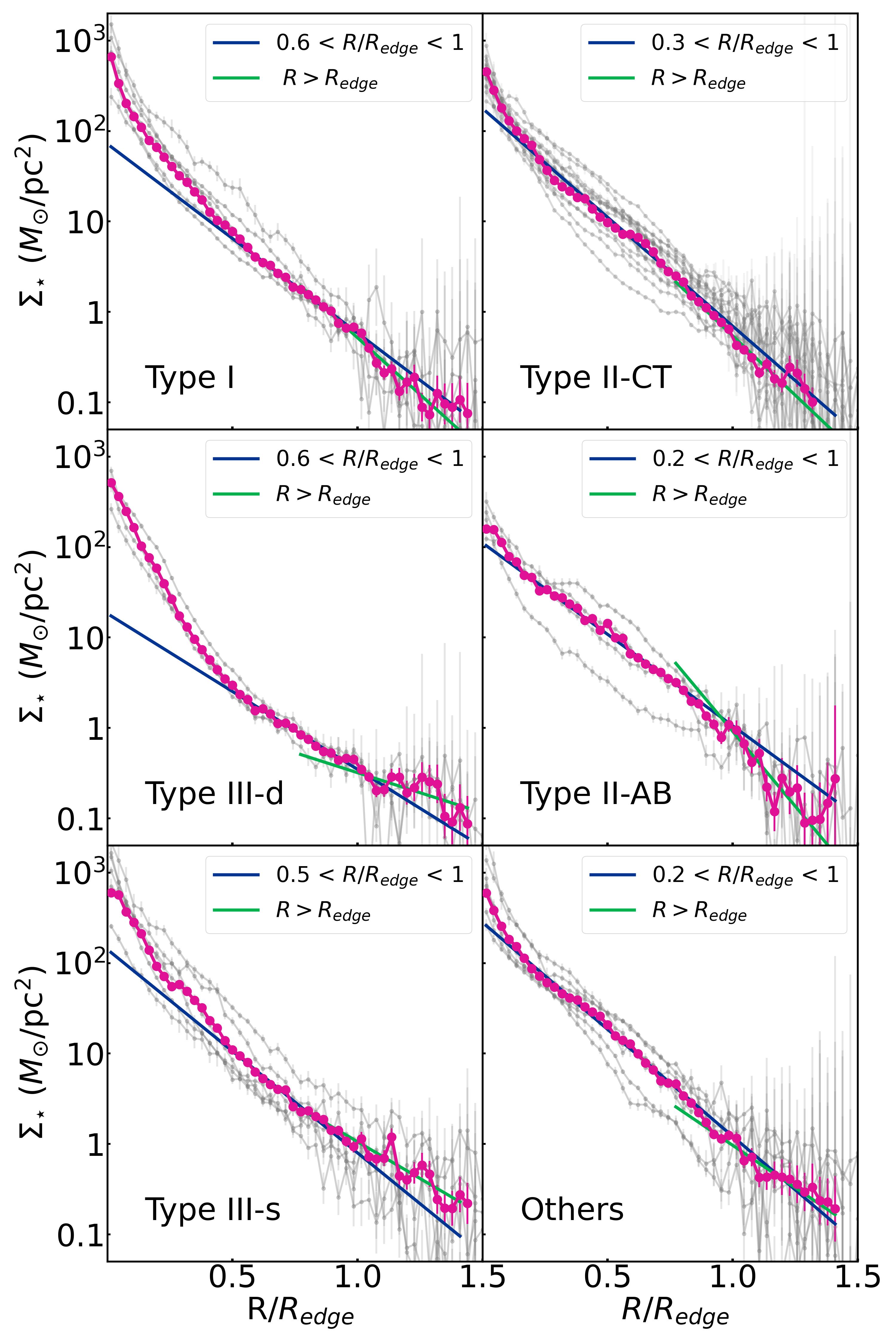}
    \caption{Stellar surface mass density profiles of the galaxies in our sample (grey lines), split into their different Types according to their surface brightness profiles (see Section \ref{classification} for details). The profiles have been normalized in the radial direction to the location of their edges \citep[according to][]{Chamba_2022}. In each panel, the average profile is shown in magenta. The blue and green lines correspond to exponential fits to the weighted average profiles inside and outside R$_{edge}$. The spatial range used to fit the inner exponential is dictated by the excess of stellar mass produced by the bulge region.}
    \label{massprofiles}
\end{figure}

\begin{figure}
    \centering
    \includegraphics[width=1\linewidth, height=1.5\linewidth]{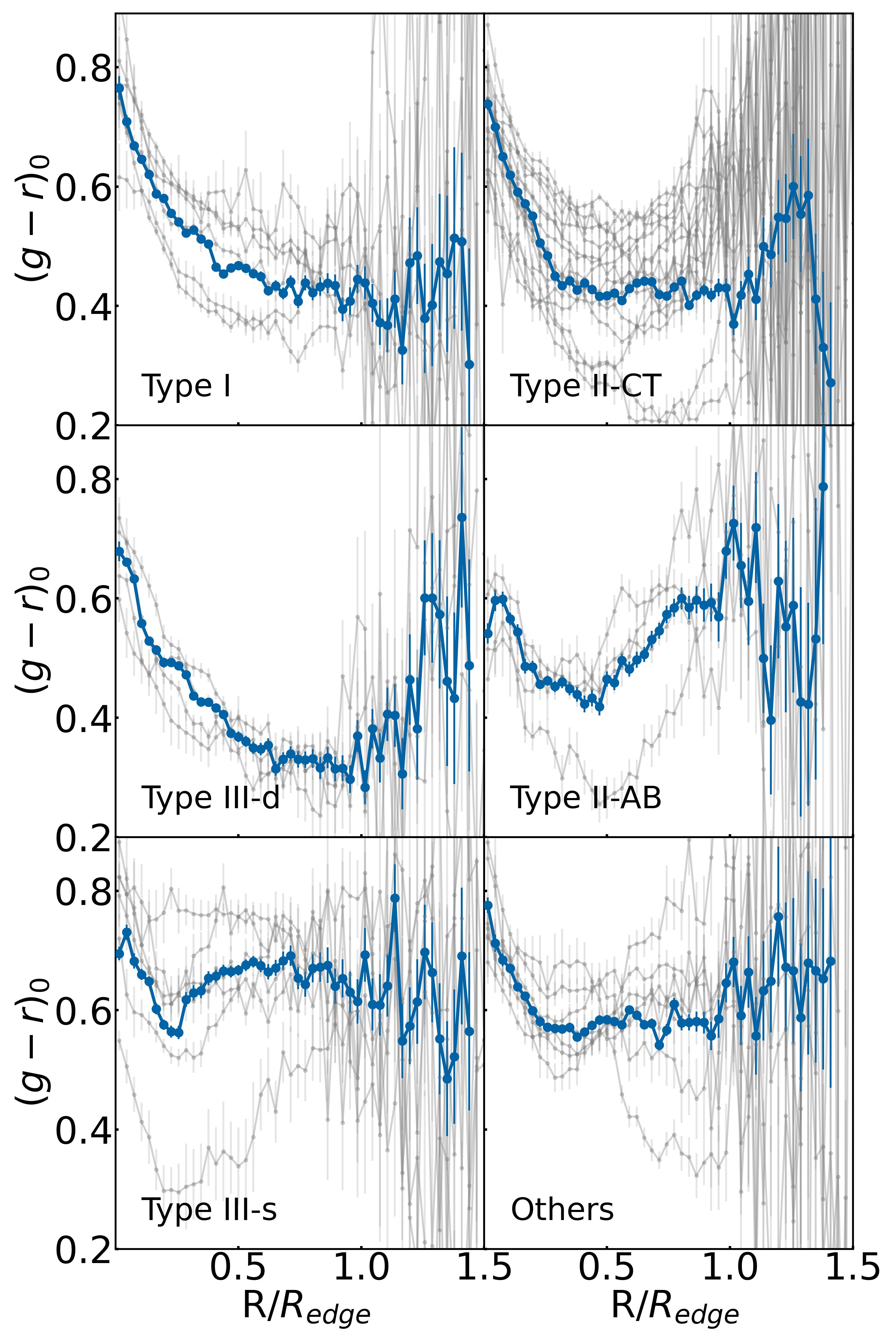}
    \caption{Similar to Fig. \ref{massprofiles} but showing the radial $(g-r)_0$ extinction corrected color profiles of the different galaxy Types and their weighted average. }
    \label{colorprofiles}
\end{figure}

 We have grouped the stellar surface mass density and color profiles of all galaxies into their different families. This is shown in Fig. \ref{massprofiles} and Fig. \ref{colorprofiles}. The radial range of the profiles has been normalized by the $R_{edge}$ of the galaxy found in \citet{Chamba_2022}. To build the average profiles at each radial distance, the uncertainties of the individual profiles were considered. Those with smaller error bars are given more weight than those with larger ones. Thus, the average profiles are the weighted mean of the individual galaxies. Explicitly, we have applied the following equation:

\begin{equation}
    <\Sigma_*(R)>=\frac{\sum\limits_{i=1}^{n}w_i(R)\Sigma_{*,i}(R)}{\sum\limits_{i=1}^{n}w_i(R)}
\end{equation}

where $w_i(R)=(\sigma_{min}(R)/\sigma_{i}(R))^2$. $\sigma_{min}(R)$ is the lower error bar of the profiles combined at given R. The new error bar at a given R of the weighted mean profiles is given by:

\begin{equation}
    \sigma^2_{ave}(R)=\frac{1}{\sum\limits_{i=1}^{n}1/\sigma^2_{i}(R)}
\end{equation}

The same weighted scheme applies to the radial color profiles shown in Fig. \ref{colorprofiles}.

As mentioned in the Introduction, we use stellar mass surface density to explore the presence of edges in galaxies because they are less susceptible to internal substructures, such as star-forming regions or spiral arms. The fact that we are averaging over a number of profiles also contributes to the smoothness of the average profiles shown in Fig. \ref{massprofiles}. This facilitates the identification of an edge, if present. In our work, the most outward change in the slope of the stellar mass profiles is what we identify as the edge of the galaxy \citep[see an in-depth discussion in][]{2025A&A...700A..91G}. In all the cases, the average $g$-band surface brightness at which R$_{edge}$ is located is around 2-3 magnitudes brighter than the limiting surface brightness of the data \citep{Chamba_2022}.

To quantify the potential presence of an edge, we perform an exponential fit on the stellar mass profile, both inside and outside the R$_{edge}$ location provided in the catalog by \cite{Chamba_2022}. The use of an exponential shape beyond the edge of the galaxy to characterize the outer disc is supported by findings from other spiral galaxies using ultra deep data, for both low \citep[see e.g.][]{trujillo2021,2025A&A...700A..91G,junais2025} and high inclinations \citep[see e.g.][]{Trujillo2016,martinezlombilla2019}. We define an exponential function in Python and fit it to the data using the \texttt{curve\_fit} from the \texttt{scipy.optimize} package to extract the best-fit parameters. The exponential fits are parameterized as follows:

\begin{equation}
    \Sigma(R)=\Sigma(0)e^{-R/h}
\end{equation}

An edge in the stellar mass profile is suggested when the ratio of the inner scale length, h$_{in}$, to the outer scale length, h$_{out}$, is greater than one. The radial range used in the innermost fitting is shown on the panels of Fig. \ref{massprofiles} and was chosen depending on the influence of the excess of light in the central part of the galaxy. For Type I, II-CT, and II-AB profiles, we can see hints of a downward slope bend when the radial distance reaches R$_{edge}$. This visual impression aligns with the  h$_{in}$ and h$_{out}$  values provided in Table \ref{scalelenghts}. The scale lengths are given in units of R$_{edge}$. We observe the opposite behavior in the other profile Types. The outer slope of the average stellar mass profile is greater than the inner slope.

The way we have combined the profiles above implicitly assumes that all stellar mass profiles within a given profile Type are different realizations of the same type of object. While this is a reasonable assumption, given that we are using objects of each Type with similar total stellar masses and similar individual and radial color profiles, it is certainly an oversimplification of reality. Therefore, to test the robustness of our scale length and axis ratio measurements, we computed these quantities for each profile, then calculated the resulting average values, i.e. $<$h$_{in}$$>$, $<$h$_{out}$$>$ and $<$h$_{in}$/h$_{out}$$>$. The results are shown in Table \ref{scalelenghts}. 

The comparison between the two methods shows that the average inner scale length is well-determined with low uncertainty, as this part of the galaxies has a higher signal-to-noise ratio. As expected, the scale length beyond the edge of the galaxy is more uncertain. Nonetheless, the estimation of this quantity agrees within the error bars for all profile Types among the two methodologies. Finally, the ratios of the two scale lengths are also in agreement (within the error bars) between the two methods. Overall, this test shows that the results presented here are robust to the analysis used to estimate them. The method using individual profiles has larger uncertainties in the outermost region, but these uncertainties are probably more realistic than those using an average profile because they take into account the intrinsic variability of all galaxies within a given Type.

\begin{table*}[]
\caption{Scale-length values (in units of R$_{edge}$) for the exponential fits inside (h$_{in}$) and outside (h$_{out}$) the R$_{edge}$.}
\begin{tabular}{|c|c|c|c|c|c|c|}
\hline
\multicolumn{1}{|c|}{Type} & I & II-CT & II-AB & III-d & III-s & Others \\ \hline
$\mu_g$(R$_{edge}$) (mag/arcsec$^2$) & $26.44\pm0.014$ & $26.40\pm0.01$ & $27.29\pm0.03$ & $26.57\pm0.01$ & $26.85\pm0.02$ & $26.94\pm0.02$ \\ \hline
\multicolumn{7}{|c|}{\textbf{Method I: Weighted Average Profiles}} \\ \hline
h$_{in}$  (R$_{edge}$)   & $0.20\pm0.01$  & $0.19\pm0.01$ & $0.21\pm0.01$  & $0.25\pm0.01$  & $0.19\pm0.01$  & $0.18\pm0.01$   \\ \hline
h$_{out}$ (R$_{edge}$)   & $0.17\pm0.02$   & $0.16\pm0.01$  & $0.13\pm0.01$   & $0.49\pm0.01$   & $0.26\pm0.04$   & $0.23\pm0.03$    \\ \hline
h$_{in}$/h$_{out}$       & $1.18\pm0.14$  & $1.19 \pm0.09$ & $1.62 \pm0.14$ & $0.51 \pm0.02$ & $0.73 \pm0.13$ & $0.78 \pm0.10$  \\ \hline
\multicolumn{7}{|c|}{\textbf{Method II: Individual Profiles}} \\ \hline
$<$h$_{in}$$>$  (R$_{edge}$)   & $0.19\pm 0.02$  &  $0.19\pm 0.01$ &  $0.20\pm 0.01$  &  $0.24\pm0.02 $  &  $0.19\pm 0.01$  &  $0.17\pm 0.01$  \\ \hline
$<$h$_{out}$$>$ (R$_{edge}$)   & $0.11\pm 0.06$  &  $0.12\pm 0.01$ &  $0.14\pm 0.03$  &  $0.34\pm 0.24$  &  $0.35\pm 0.07$  &  $0.26\pm 0.30$    \\ \hline
$<$h$_{in}$/h$_{out}$$>$      & $1.45\pm 0.22$  &  $1.51\pm0.10 $ &  $1.34\pm0.30$  &  $0.70\pm 0.25$  &  $0.47\pm 0.20$  &  $0.81\pm 0.20$  \\ \hline

\end{tabular}
\tablefoot{R$_{edge}$ location provided in the catalog by \citet{Chamba_2022}. The parameters were estimated using two different approaches: a) using a weighted average mass profile for each Type and b) using individual stellar mass profiles and calculating the average values afterward.  For Method I, we computed the errors of the ratio ($h_{in}/h_{out}$) using standard error propagation. For Method II, we estimated the uncertainty in the average by dividing the standard deviation by the square root of the number of galaxies of that type. The table also shows the average surface brightness (in the $g$-band) of the different profile types at R$_{edge}$.}
\label{scalelenghts}
\end{table*}

For all profile Types, we find that the inner scale length is approximately five times smaller than R$_{edge}$. This result is in good agreement with that found by \citet{Chamba_2022} when comparing with the global effective radius for MW-like galaxies (i.e., R$_{edge}$/R$_e$$\sim$3.2). Given that, for exponential profiles, R$_e$/h$\sim$1.7, we can expect R$_{edge}$/h$_{in}$$\sim$5.

While the absence of a measurable edge in the stellar mass profile of galaxies of Types III-s and 'Others', which is a combination in this work of Types  II.o-OLR, and II.i is expected, the lack of an edge in Type III-d is surprising. In the case of Type III-s, the excess stellar mass beyond the in situ star formation region of the galaxy can be easily explained by ongoing stellar mass accretion. In the case of the 'Others' category, it is also easy to see that we are dealing with a mixture of different profiles, which is why there is no clear edge. However, it is not obvious why there is no definite edge in Type III-d. The surprise is even greater when we take a detailed look at the color profiles given in Fig. \ref{colorprofiles}. Types I, II share a reddening after the edge together with Type III-d. This is a typical feature associated with the end of the in situ star formation region. For these reasons, we have decided to explore the four galaxies in our sample with III-d profiles in more detail.

\begin{figure*}
    \centering
    \includegraphics[width=\linewidth, height=1.2\linewidth]{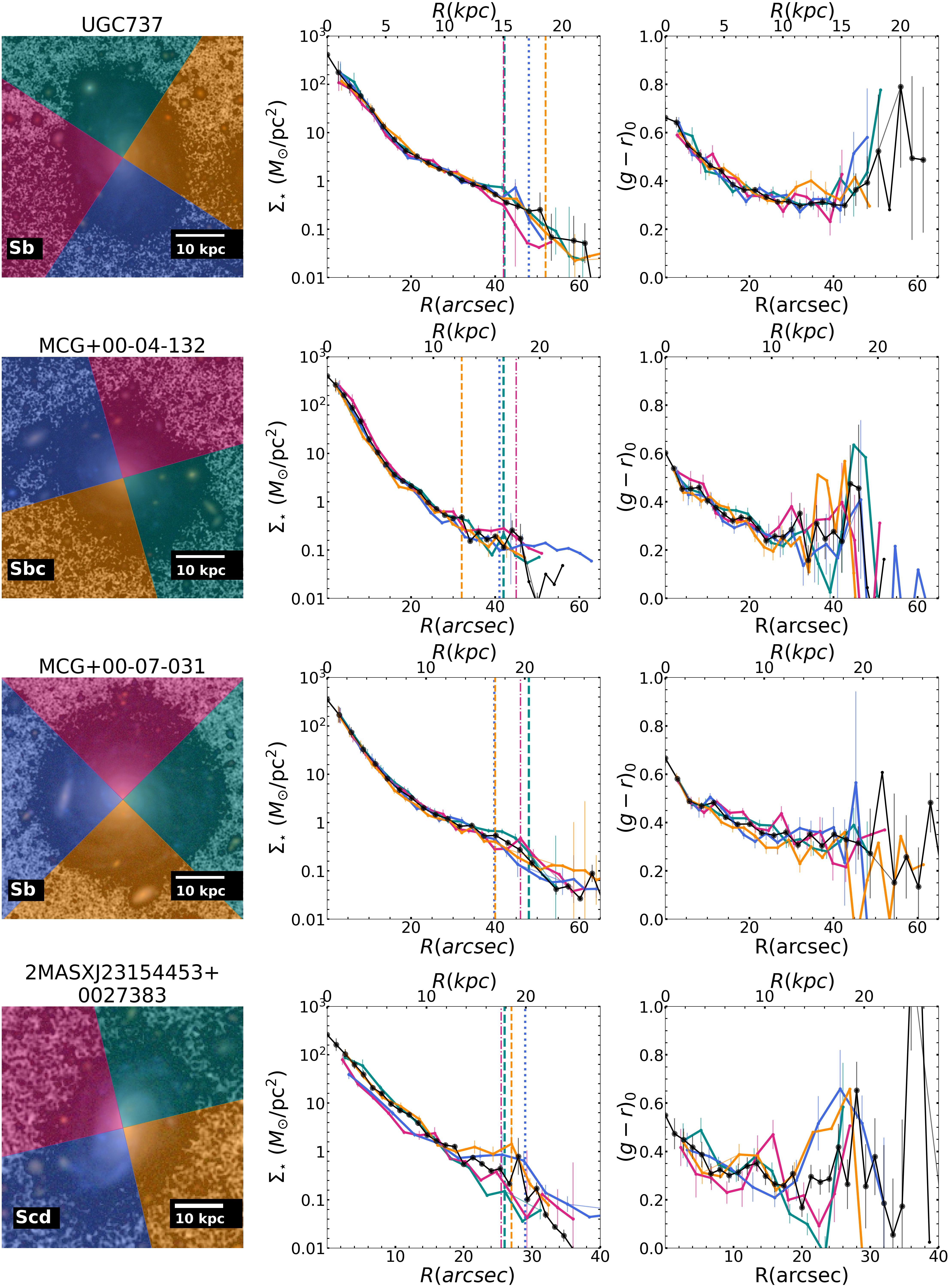}
    \caption{Color image and radial profiles of Type III-d galaxies using wedges (as indicated in the colored region). The left column shows a composite color image of the galaxy using the $g$, $r$, and $i$ Sloan bands.  The middle column displays the stellar mass profiles along the different wedges. The vertical dashed lines indicate the location of the edge along each wedge direction. The black profile is the elliptically averaged stellar mass density profile. The right column shows the Galactic extinction corrected color wedge profiles of the galaxy together with the elliptically averaged one (black line).}
    \label{galaxywedges}
\end{figure*}

Figures in appendix \ref{profiles} show images and surface brightness, $(g-r)_0$ color, and stellar mass density radial profiles (using ellipses) of the galaxies in our sample. Type III-d galaxies have larger disk asymmetries than the other galaxies in the sample. MCG+00-07-031 and 2MASXJ23154453+0027383 are the most obvious cases, visually looking at them. In these cases, using an elliptical average to obtain the surface brightness profiles can obscure the objects' features. In particular, finding the edge location is very sensitive to asymmetries in the outer disks using this approach \citep[see e.g.][]{junais2025}. To overcome this difficulty, we obtained the surface brightness profiles of these objects at specific angular directions using \texttt{astscript-radial-profile} as described in \cite {astscript-radial-profile}. This wedge strategy is illustrated in Fig. \ref{galaxywedges}. The direction of the wedges were selected based on minimizing background contamination and ensuring representativeness of relatively symmetric areas of the galaxy.

\begin{figure*}
    \centering
    \includegraphics[width=1\linewidth, height=0.5\linewidth]{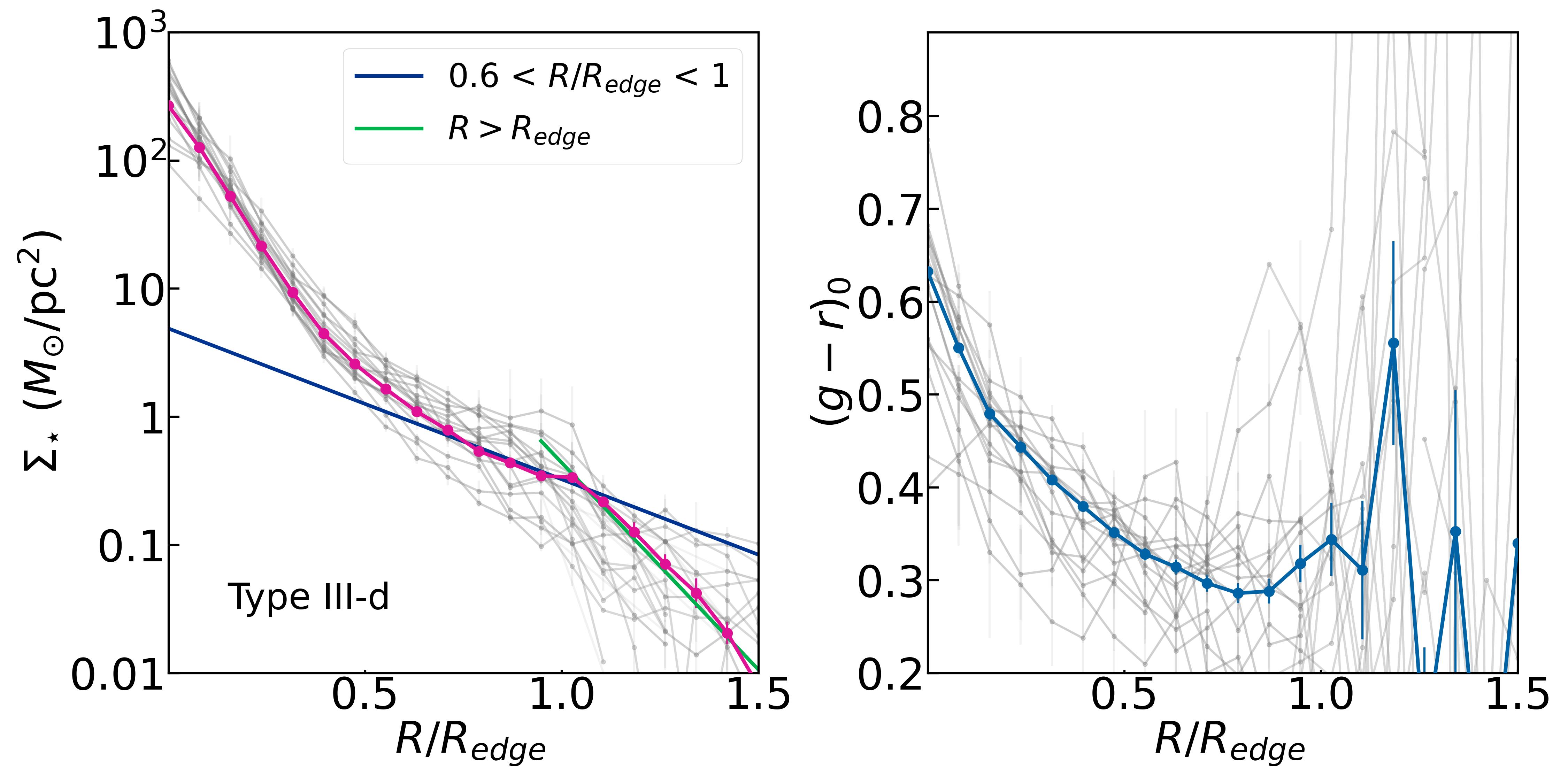}
    \caption{Similar to Fig. \ref{massprofiles} and Fig. \ref{colorprofiles} but showing the stellar mass density and $(g-r)_0$ color profiles of the Type III-d galaxies using wedges.}
    \label{masscolorwedges}
\end{figure*}

When doing the above analysis, we find that the elliptically averaged profile makes the finding of the edges more difficult than in the case of the individual profiles following the wedge directions. We have repeated the analysis of Type III-d galaxies, combining all the stellar mass and color profiles extracted using the wedge technique. This is shown in Fig. \ref{masscolorwedges}. Contrary to what we observe using the elliptically averaged profiles, the wedge technique shows a very clear edge for the Type III-d galaxies. In fact, we get ratio h$_{in}$/h$_{out}$=2.74$\pm$0.45 with h$_{in}$=0.326$\pm$0.06 and h$_{out}$=0.13$\pm$0.006.

After this analysis, we conclude that Type III-d galaxies do have a well-defined edge when viewed from a specific direction. This confirms that the reason we cannot see the edge using ellipses is due to the asymmetry of these galaxies' disks. In short, there is strong evidence that Types I, II-CT, and III-d galaxies exhibit edges in their stellar mass profiles, despite their varying surface brightness profiles.

\section{Discussion and Conclusions}
\label{summary}

The discovery that disk galaxy surface brightness profiles can be categorized into three Types naturally raises the question of why this variety emerged. To shed light on this issue, this study examines a carefully selected group of disk galaxies using deep images obtained from the IAC Stripe82 Legacy Project \citep{2016MNRAS.456.1359F,2018RNAAS...2..144R}. Having deeper images than those used in seminal works helps us explore the outermost regions of galaxies and obtain stellar mass profiles through their color distributions, which are more reliable due to the increased signal-to-noise ratio.

The first finding of our study, based on the analysis of stellar mass density profiles, is that disk galaxies of Types I, II-CT, and III-d exhibit an edge. These edges are found at surface brightness levels ($\mu_g$(R$_{edge}$)$\sim$26.5 mag/arcsec$^2$)  which are very close to the limiting surface brightness of typical images with depth, such as those from the Sloan Digital Sky Survey (SDSS) twenty years ago. For this reason, they were practically invisible to those seminal studies. In all galaxy Types, the edge seems to indicate the end of in situ star formation in the disk. This is because, in addition to having similar surface brightnesses, they also have similar average stellar surface densities at the edge ($\Sigma$(R$_{edge}$)$\sim$0.5-1 M$_{\odot}$/pc$^2$) across the different categories. The stellar populations beyond the edge are redder, reinforcing the idea that in situ star formation ends at the edge. While this could support the hypothesis that stars beyond the edge of galaxies result from stellar migration, other plausible scenarios, such as gas depletion and/or disk flaring due to interactions, could also play an important role.

Although this point is not discussed in detail in this paper, our findings from studying the edges of Type I, II-CT, and III-d galaxies suggest that the next generation of images \citep[see e.g.][]{2019lsst, LIGHTSs,2024Zaritskydennislights}, which are deeper than those analyzed in this paper, will require us to change the traditional notation of Types that we have used until now. In the future, deeper surface brightness profiles of previous Type galaxies will show a break in the outer part of the profiles, corresponding to the edge feature. Thus, the different Types will probably be classified as I+II,  II+II, or III-d+II, with the final II indicating the presence of the galaxy's edge.

Another important aspect to discuss is the following:  We have taken disk galaxies with very similar stellar masses, and all of the Type I, II-CT, and III-d galaxies have edges in their mass profiles with similar properties \citep[i.e., similar radial location and stellar mass surface density, see e.g. Figure 5 in ][]{Chamba_2022}. Why, then, do their surface brightness profiles show differences in internal structure? In short, what is the ultimate physical reason for the disparity in their surface brightness profiles? The answer seems to be related to the galaxies' morphological type. This result was previously suggested by \citet{Pohlen_2006}. In that study, Type II-CT galaxies were more prevalent among later types, while Type III-d galaxies were more abundant among earlier types\footnote{We found that nine out of sixteen of our Type II-CT galaxies have a morphological type of T=5 (i.e., Sc). There is no obvious clustering among Type III-d galaxies; however, two out of four of these galaxies have a morphological type of T=3 (i.e., Sb). Nevertheless, the numbers (particularly in the case of Type III-d galaxies) are admittedly small, and after running a KS test, we cannot reject the null hypothesis that the two samples belong to the same parent sample (p-value = 0.21).}. The shape of our galaxies' stellar mass profiles reinforces this idea. For instance, Type II-CT galaxies have less prominent bulge regions compared to their disk (exponential) parts. The opposite is true for Type III-d galaxies. This is also evident in the color stacked profiles, where the flat region with the bluest color is larger in Type II-CT ($0.3 <$ $R/R_{edge}$ $< 1$) galaxies than in Type III-d ($0.5 <$ $R/R_{edge}$ $< 1$). In addition to the extension of the exponential region among the different Types, the color of the disk region is also a bit different. Types I and II-CT have similar blue colors around $(g-r)_0$$\sim$0.4 while Type III-d has colors reaching values around 0.3 mag.  This, together with the fact that the exponential region ($\Delta$ $R_{\mathrm{exp}}$) of Type III-d galaxies has a very low stellar mass density ($\Sigma$($\Delta$ $R_{\mathrm{exp}}$)$\sim$0.5 M$_{\odot}$/pc$^2$), leads us to propose that Type III-d galaxies have the newest disk structures. Future studies with a larger sample of galaxies, particularly Type III-d galaxies, will conclusively determine whether the differences between Types II-CT and III-d identified here are robust or the result of statistical fluctuations.

Finally, the differences between Types I and II-CT are subtle and may relate to the bulge's significance compared to the disk. Type I galaxies have larger central regions ($0< R/R_{edge} <0.25$) than Type II-CT galaxies ($0< R/R_{edge} <0.1$). Could this be connected to the presence of a break in the surface brightness profile of Type II-CT galaxies? Does a more prominent bulge prevent the disk from resonating, thereby enhancing star formation at the location of the galaxy's break? Answering these questions is beyond the scope of this work, and only simulations built specifically to address them will be able to provide solutions.

\begin{acknowledgements}

We thank the anonymous referee for carefully reading the manuscript and providing suggestions that helped improve its clarity and strength. SR and FB acknowledge support from the GEELSBE2 project with reference PID2023-150393NB-I00 funded by MCIU/AEI/10.13039/501100011033 and the FSE+, and also the Consolidación Investigadora IGADLE project with reference CNS2024-154572. SR and FB also acknowledge support from the project PID2020-116188GA-I00, funded by MICIU/AEI/10.13039/501100011033. SR and FB gratefully acknowledge financial support of the Department of Education, Junta de Castilla y León, and FEDER Funds (Reference: CLU-2023-1-05). IT acknowledges support from the State Research Agency (AEI-MCINN) of the Spanish Ministry of Science and Innovation under the grant PID2022-140869NB-I00 and IAC project P/302302, financed by the Ministry of Science and Innovation, through the State Budget and by the Canary Islands Department of Economy, Knowledge, and Employment, through the Regional Budget of the Autonomous Community. This research also acknowledges support from the European Union through the following grants: "UNDARK" and "Excellence in Galaxies - Twinning the IAC" of the EU Horizon Europe Widening Actions programs (project numbers 101159929 and 101158446). Funding for this work/research was provided by the European Union (MSCA EDUCADO, GA 101119830). Views and opinions expressed are however those of the author(s) only and do not necessarily reflect those of the European Union or European Research Executive Agency (REA). Neither the European Union nor the granting authority can be held responsible for them. GG and IRC acknowledge support from the PID2022-140869NB-I00 grant from the Spanish Ministry of Science and Innovation. 
\end{acknowledgements}

\bibliographystyle{aa}
\bibliography{authors.bib}

\begin{thebibliography}{53}
\expandafter\ifx\csname natexlab\endcsname\relax\def\natexlab#1{#1}\fi

\bibitem[{{Abazajian} {et~al.}(2009){Abazajian}, {Adelman-McCarthy}, {Ag{\"u}eros}, {Allam}, {Allende Prieto}, {An}, {Anderson}, {Anderson}, {Annis}, {Bahcall}, {Bailer-Jones}, {Barentine}, {Bassett}, {Becker}, {Beers}, {Bell}, {Belokurov}, {Berlind}, {Berman}, {Bernardi}, {Bickerton}, {Bizyaev}, {Blakeslee}, {Blanton}, {Bochanski}, {Boroski}, {Brewington}, {Brinchmann}, {Brinkmann}, {Brunner}, {Budav{\'a}ri}, {Carey}, {Carliles}, {Carr}, {Castander}, {Cinabro}, {Connolly}, {Csabai}, {Cunha}, {Czarapata}, {Davenport}, {de Haas}, {Dilday}, {Doi}, {Eisenstein}, {Evans}, {Evans}, {Fan}, {Friedman}, {Frieman}, {Fukugita}, {G{\"a}nsicke}, {Gates}, {Gillespie}, {Gilmore}, {Gonzalez}, {Gonzalez}, {Grebel}, {Gunn}, {Gy{\"o}ry}, {Hall}, {Harding}, {Harris}, {Harvanek}, {Hawley}, {Hayes}, {Heckman}, {Hendry}, {Hennessy}, {Hindsley}, {Hoblitt}, {Hogan}, {Hogg}, {Holtzman}, {Hyde}, {Ichikawa}, {Ichikawa}, {Im}, {Ivezi{\'c}}, {Jester}, {Jiang}, {Johnson}, {Jorgensen}, {Juri{\'c}}, {Kent}, {Kessler}, {Kleinman}, {Knapp},
  {Konishi}, {Kron}, {Krzesinski}, {Kuropatkin}, {Lampeitl}, {Lebedeva}, {Lee}, {Lee}, {French Leger}, {L{\'e}pine}, {Li}, {Lima}, {Lin}, {Long}, {Loomis}, {Loveday}, {Lupton}, {Magnier}, {Malanushenko}, {Malanushenko}, {Mandelbaum}, {Margon}, {Marriner}, {Mart{\'\i}nez-Delgado}, {Matsubara}, {McGehee}, {McKay}, {Meiksin}, {Morrison}, {Mullally}, {Munn}, {Murphy}, {Nash}, {Nebot}, {Neilsen}, {Newberg}, {Newman}, {Nichol}, {Nicinski}, {Nieto-Santisteban}, {Nitta}, {Okamura}, {Oravetz}, {Ostriker}, {Owen}, {Padmanabhan}, {Pan}, {Park}, {Pauls}, {Peoples}, {Percival}, {Pier}, {Pope}, {Pourbaix}, {Price}, {Purger}, {Quinn}, {Raddick}, {Re Fiorentin}, {Richards}, {Richmond}, {Riess}, {Rix}, {Rockosi}, {Sako}, {Schlegel}, {Schneider}, {Scholz}, {Schreiber}, {Schwope}, {Seljak}, {Sesar}, {Sheldon}, {Shimasaku}, {Sibley}, {Simmons}, {Sivarani}, {Allyn Smith}, {Smith}, {Smol{\v{c}}i{\'c}}, {Snedden}, {Stebbins}, {Steinmetz}, {Stoughton}, {Strauss}, {SubbaRao}, {Suto}, {Szalay}, {Szapudi}, {Szkody}, {Tanaka},
  {Tegmark}, {Teodoro}, {Thakar}, {Tremonti}, {Tucker}, {Uomoto}, {Vanden Berk}, {Vandenberg}, {Vidrih}, {Vogeley}, {Voges}, {Vogt}, {Wadadekar}, {Watters}, {Weinberg}, {West}, {White}, {Wilhite}, {Wonders}, {Yanny}, {Yocum}, {York}, {Zehavi}, {Zibetti}, \& {Zucker}}]{2009ApJS..182..543A}
{Abazajian}, K.~N., {Adelman-McCarthy}, J.~K., {Ag{\"u}eros}, M.~A., {et~al.} 2009, \apjs, 182, 543

\bibitem[{{Akhlaghi} \& {Ichikawa}(2015)}]{gnuastro}
{Akhlaghi}, M. \& {Ichikawa}, T. 2015, ApJS, 220, 1

\bibitem[{{Azzollini} {et~al.}(2008){Azzollini}, {Trujillo}, \& {Beckman}}]{Azzollini2008}
{Azzollini}, R., {Trujillo}, I., \& {Beckman}, J.~E. 2008, \apjl, 679, L69

\bibitem[{{Bakos} \& {Trujillo}(2012)}]{bakos2012}
{Bakos}, J. \& {Trujillo}, I. 2012, arXiv e-prints, arXiv:1204.3082

\bibitem[{{Bakos} {et~al.}(2008){Bakos}, {Trujillo}, \& {Pohlen}}]{Bakos2008}
{Bakos}, J., {Trujillo}, I., \& {Pohlen}, M. 2008, \apjl, 683, L103

\bibitem[{{Buitrago} \& {Trujillo}(2024)}]{Buitrago2023}
{Buitrago}, F. \& {Trujillo}, I. 2024, \aap, 682, A110

\bibitem[{{Capaccioli} {et~al.}(2015){Capaccioli}, {Spavone}, {Grado}, {Iodice}, {Limatola}, {Napolitano}, {Cantiello}, {Paolillo}, {Romanowsky}, {Forbes}, {Puzia}, {Raimondo}, \& {Schipani}}]{2015A&A...581A..10C}
{Capaccioli}, M., {Spavone}, M., {Grado}, A., {et~al.} 2015, \aap, 581, A10

\bibitem[{{Chamba} {et~al.}(2022){Chamba}, {Trujillo}, \& {Knapen}}]{Chamba_2022}
{Chamba}, N., {Trujillo}, I., \& {Knapen}, J.~H. 2022, \aap, 667, A87

\bibitem[{{Debattista} {et~al.}(2006){Debattista}, {Mayer}, {Carollo}, {Moore}, {Wadsley}, \& {Quinn}}]{Debattista2006}
{Debattista}, V.~P., {Mayer}, L., {Carollo}, C.~M., {et~al.} 2006, \apj, 645, 209

\bibitem[{{Duc} {et~al.}(2015){Duc}, {Cuillandre}, {Karabal}, {Cappellari}, {Alatalo}, {Blitz}, {Bournaud}, {Bureau}, {Crocker}, {Davies}, {Davis}, {de Zeeuw}, {Emsellem}, {Khochfar}, {Krajnovi{\'c}}, {Kuntschner}, {McDermid}, {Michel-Dansac}, {Morganti}, {Naab}, {Oosterloo}, {Paudel}, {Sarzi}, {Scott}, {Serra}, {Weijmans}, \& {Young}}]{2015MNRAS.446..120D}
{Duc}, P.-A., {Cuillandre}, J.-C., {Karabal}, E., {et~al.} 2015, \mnras, 446, 120

\bibitem[{{Erwin} {et~al.}(2005){Erwin}, {Beckman}, \& {Pohlen}}]{Erwin2005}
{Erwin}, P., {Beckman}, J.~E., \& {Pohlen}, M. 2005, \apjl, 626, L81

\bibitem[{{Ferrarese} {et~al.}(2012){Ferrarese}, {C{\^o}t{\'e}}, {Cuillandre}, {Gwyn}, {Peng}, {MacArthur}, {Duc}, {Boselli}, {Mei}, {Erben}, {McConnachie}, {Durrell}, {Mihos}, {Jord{\'a}n}, {Lan{\c{c}}on}, {Puzia}, {Emsellem}, {Balogh}, {Blakeslee}, {van Waerbeke}, {Gavazzi}, {Vollmer}, {Kavelaars}, {Woods}, {Ball}, {Boissier}, {Courteau}, {Ferriere}, {Gavazzi}, {Hildebrandt}, {Hudelot}, {Huertas-Company}, {Liu}, {McLaughlin}, {Mellier}, {Milkeraitis}, {Schade}, {Balkowski}, {Bournaud}, {Carlberg}, {Chapman}, {Hoekstra}, {Peng}, {Sawicki}, {Simard}, {Taylor}, {Tully}, {van Driel}, {Wilson}, {Burdullis}, {Mahoney}, \& {Manset}}]{2012ApJS..200....4F}
{Ferrarese}, L., {C{\^o}t{\'e}}, P., {Cuillandre}, J.-C., {et~al.} 2012, \apjs, 200, 4

\bibitem[{{Fliri} \& {Trujillo}(2016)}]{2016MNRAS.456.1359F}
{Fliri}, J. \& {Trujillo}, I. 2016, \mnras, 456, 1359

\bibitem[{{Freeman}(1970)}]{1970ApJ...160..811F}
{Freeman}, K.~C. 1970, \apj, 160, 811

\bibitem[{{Golini} {et~al.}(2024){Golini}, {Montes}, {Carrasco}, {Rom{\'a}n}, \& {Trujillo}}]{2024A&A...684A..99G}
{Golini}, G., {Montes}, M., {Carrasco}, E.~R., {Rom{\'a}n}, J., \& {Trujillo}, I. 2024, \aap, 684, A99

\bibitem[{{Golini} {et~al.}(2025){Golini}, {Trujillo}, {Zaritsky}, {Montes}, {Infante-Sainz}, {Martin}, {Chamba}, {Ruiz Cejudo}, {Asensio Ramos}, {Chuang}, {D'Onofrio}, {Eskandarlou}, {Zahra Hosseini-ShahiSavandi}, {Kaboud}, {Marrero de la Rosa}, {Ngoc Le}, {Raji}, {Rom{\'a}n}, {Sedighi}, {Sharbaf}, {Donnerstein}, \& {Guerra Arencibia}}]{2025A&A...700A..91G}
{Golini}, G., {Trujillo}, I., {Zaritsky}, D., {et~al.} 2025, \aap, 700, A91

\bibitem[{{Haigh} {et~al.}(2021){Haigh}, {Chamba}, {Venhola}, {Peletier}, {Doorenbos}, {Watkins}, \& {Wilkinson}}]{2021A&A...645A.107H}
{Haigh}, C., {Chamba}, N., {Venhola}, A., {et~al.} 2021, \aap, 645, A107

\bibitem[{{Hood} {et~al.}(2018){Hood}, {Kannappan}, {Stark}, {Dell'Antonio}, {Moffett}, {Eckert}, {Norris}, \& {Hendel}}]{2018ApJ...857..144H}
{Hood}, C.~E., {Kannappan}, S.~J., {Stark}, D.~V., {et~al.} 2018, \apj, 857, 144

\bibitem[{{Huang} {et~al.}(2018){Huang}, {Leauthaud}, {Greene}, {Bundy}, {Lin}, {Tanaka}, {Mandelbaum}, {Miyazaki}, \& {Komiyama}}]{2018MNRAS.480..521H}
{Huang}, S., {Leauthaud}, A., {Greene}, J., {et~al.} 2018, \mnras, 480, 521

\bibitem[{{Infante-Sainz} \& {Akhlaghi}(2024)}]{astscript-color-faint-gray}
{Infante-Sainz}, R. \& {Akhlaghi}, M. 2024, Research Notes of the American Astronomical Society, 8, 10

\bibitem[{{Infante-Sainz} {et~al.}(2024){Infante-Sainz}, {Akhlaghi}, \& {Eskandarlou}}]{astscript-radial-profile}
{Infante-Sainz}, R., {Akhlaghi}, M., \& {Eskandarlou}, S. 2024, Research Notes of the American Astronomical Society, 8, 22

\bibitem[{Infante-Sainz {et~al.}(2020)Infante-Sainz, Trujillo, \& Rom{\'a}n}]{infante2020}
Infante-Sainz, R., Trujillo, I., \& Rom{\'a}n, J. 2020, Monthly Notices of the Royal Astronomical Society, 491, 5317

\bibitem[{{Ivezi{\'c}} {et~al.}(2019){Ivezi{\'c}}, {Kahn}, {Tyson}, {Abel}, {Acosta}, {Allsman}, {Alonso}, {AlSayyad}, {Anderson}, {Andrew}, {Angel}, {Angeli}, {Ansari}, {Antilogus}, {Araujo}, {Armstrong}, {Arndt}, {Astier}, {Aubourg}, {Auza}, {Axelrod}, {Bard}, {Barr}, {Barrau}, {Bartlett}, {Bauer}, {Bauman}, {Baumont}, {Bechtol}, {Bechtol}, {Becker}, {Becla}, {Beldica}, {Bellavia}, {Bianco}, {Biswas}, {Blanc}, {Blazek}, {Blandford}, {Bloom}, {Bogart}, {Bond}, {Booth}, {Borgland}, {Borne}, {Bosch}, {Boutigny}, {Brackett}, {Bradshaw}, {Brandt}, {Brown}, {Bullock}, {Burchat}, {Burke}, {Cagnoli}, {Calabrese}, {Callahan}, {Callen}, {Carlin}, {Carlson}, {Chandrasekharan}, {Charles-Emerson}, {Chesley}, {Cheu}, {Chiang}, {Chiang}, {Chirino}, {Chow}, {Ciardi}, {Claver}, {Cohen-Tanugi}, {Cockrum}, {Coles}, {Connolly}, {Cook}, {Cooray}, {Covey}, {Cribbs}, {Cui}, {Cutri}, {Daly}, {Daniel}, {Daruich}, {Daubard}, {Daues}, {Dawson}, {Delgado}, {Dellapenna}, {de Peyster}, {de Val-Borro}, {Digel}, {Doherty}, {Dubois},
  {Dubois-Felsmann}, {Durech}, {Economou}, {Eifler}, {Eracleous}, {Emmons}, {Fausti Neto}, {Ferguson}, {Figueroa}, {Fisher-Levine}, {Focke}, {Foss}, {Frank}, {Freemon}, {Gangler}, {Gawiser}, {Geary}, {Gee}, {Geha}, {Gessner}, {Gibson}, {Gilmore}, {Glanzman}, {Glick}, {Goldina}, {Goldstein}, {Goodenow}, {Graham}, {Gressler}, {Gris}, {Guy}, {Guyonnet}, {Haller}, {Harris}, {Hascall}, {Haupt}, {Hernandez}, {Herrmann}, {Hileman}, {Hoblitt}, {Hodgson}, {Hogan}, {Howard}, {Huang}, {Huffer}, {Ingraham}, {Innes}, {Jacoby}, {Jain}, {Jammes}, {Jee}, {Jenness}, {Jernigan}, {Jevremovi{\'c}}, {Johns}, {Johnson}, {Johnson}, {Jones}, {Juramy-Gilles}, {Juri{\'c}}, {Kalirai}, {Kallivayalil}, {Kalmbach}, {Kantor}, {Karst}, {Kasliwal}, {Kelly}, {Kessler}, {Kinnison}, {Kirkby}, {Knox}, {Kotov}, {Krabbendam}, {Krughoff}, {Kub{\'a}nek}, {Kuczewski}, {Kulkarni}, {Ku}, {Kurita}, {Lage}, {Lambert}, {Lange}, {Langton}, {Le Guillou}, {Levine}, {Liang}, {Lim}, {Lintott}, {Long}, {Lopez}, {Lotz}, {Lupton}, {Lust}, {MacArthur}, {Mahabal},
  {Mandelbaum}, {Markiewicz}, {Marsh}, {Marshall}, {Marshall}, {May}, {McKercher}, {McQueen}, {Meyers}, {Migliore}, {Miller}, {Mills}, {Miraval}, {Moeyens}, {Moolekamp}, {Monet}, {Moniez}, {Monkewitz}, {Montgomery}, {Morrison}, {Mueller}, {Muller}, {Mu{\~n}oz Arancibia}, {Neill}, {Newbry}, {Nief}, {Nomerotski}, {Nordby}, {O'Connor}, {Oliver}, {Olivier}, {Olsen}, {O'Mullane}, {Ortiz}, {Osier}, {Owen}, {Pain}, {Palecek}, {Parejko}, {Parsons}, {Pease}, {Peterson}, {Peterson}, {Petravick}, {Libby Petrick}, {Petry}, {Pierfederici}, {Pietrowicz}, {Pike}, {Pinto}, {Plante}, {Plate}, {Plutchak}, {Price}, {Prouza}, {Radeka}, {Rajagopal}, {Rasmussen}, {Regnault}, {Reil}, {Reiss}, {Reuter}, {Ridgway}, {Riot}, {Ritz}, {Robinson}, {Roby}, {Roodman}, {Rosing}, {Roucelle}, {Rumore}, {Russo}, {Saha}, {Sassolas}, {Schalk}, {Schellart}, {Schindler}, {Schmidt}, {Schneider}, {Schneider}, {Schoening}, {Schumacher}, {Schwamb}, {Sebag}, {Selvy}, {Sembroski}, {Seppala}, {Serio}, {Serrano}, {Shaw}, {Shipsey}, {Sick}, {Silvestri},
  {Slater}, {Smith}, {Smith}, {Sobhani}, {Soldahl}, {Storrie-Lombardi}, {Stover}, {Strauss}, {Street}, {Stubbs}, {Sullivan}, {Sweeney}, {Swinbank}, {Szalay}, {Takacs}, {Tether}, {Thaler}, {Thayer}, {Thomas}, {Thornton}, {Thukral}, {Tice}, {Trilling}, {Turri}, {Van Berg}, {Vanden Berk}, {Vetter}, {Virieux}, {Vucina}, {Wahl}, {Walkowicz}, {Walsh}, {Walter}, {Wang}, {Wang}, {Warner}, {Wiecha}, {Willman}, {Winters}, {Wittman}, {Wolff}, {Wood-Vasey}, {Wu}, {Xin}, {Yoachim}, \& {Zhan}}]{2019lsst}
{Ivezi{\'c}}, {\v{Z}}., {Kahn}, S.~M., {Tyson}, J.~A., {et~al.} 2019, \apj, 873, 111

\bibitem[{{Jiang} {et~al.}(2008){Jiang}, {Fan}, {Annis}, {Becker}, {White}, {Chiu}, {Lin}, {Lupton}, {Richards}, {Strauss}, {Jester}, \& {Schneider}}]{2008AJ....135.1057J}
{Jiang}, L., {Fan}, X., {Annis}, J., {et~al.} 2008, \aj, 135, 1057

\bibitem[{{Junais} {et~al.}(2025){Junais}, {Ruiz Cejudo}, {Guerra Arencibia}, {Trujillo}, {Alarcon}, {Serra-Ricart}, {Knapen}, \& {Duc}}]{junais2025}
{Junais}, {Ruiz Cejudo}, I., {Guerra Arencibia}, S., {et~al.} 2025, \aap, 702, A136

\bibitem[{{Kelvin} {et~al.}(2014){Kelvin}, {Driver}, {Robotham}, {Taylor}, {Graham}, {Alpaslan}, {Baldry}, {Bamford}, {Bauer}, {Bland-Hawthorn}, {Brown}, {Colless}, {Conselice}, {Holwerda}, {Hopkins}, {Lara-L{\'o}pez}, {Liske}, {L{\'o}pez-S{\'a}nchez}, {Loveday}, {Norberg}, {Phillipps}, {Popescu}, {Prescott}, {Sansom}, \& {Tuffs}}]{2014MNRAS.444.1647K}
{Kelvin}, L.~S., {Driver}, S.~P., {Robotham}, A. S.~G., {et~al.} 2014, \mnras, 444, 1647

\bibitem[{{Mart{\'\i}n-Navarro} {et~al.}(2012{\natexlab{a}}){Mart{\'\i}n-Navarro}, {Bakos}, {Trujillo}, {Knapen}, {Athanassoula}, {Bosma}, {Comer{\'o}n}, {Elmegreen}, {Erroz-Ferrer}, {Gadotti}, {Gil de Paz}, {Hinz}, {Ho}, {Holwerda}, {Kim}, {Laine}, {Laurikainen}, {Men{\'e}ndez-Delmestre}, {Mizusawa}, {Mu{\~n}oz-Mateos}, {Regan}, {Salo}, {Seibert}, \& {Sheth}}]{martin-navarro2012}
{Mart{\'\i}n-Navarro}, I., {Bakos}, J., {Trujillo}, I., {et~al.} 2012{\natexlab{a}}, \mnras, 427, 1102

\bibitem[{{Mart{\'\i}n-Navarro} {et~al.}(2012{\natexlab{b}}){Mart{\'\i}n-Navarro}, {Bakos}, {Trujillo}, {Knapen}, {Athanassoula}, {Bosma}, {Comer{\'o}n}, {Elmegreen}, {Erroz-Ferrer}, {Gadotti}, {Gil de Paz}, {Hinz}, {Ho}, {Holwerda}, {Kim}, {Laine}, {Laurikainen}, {Men{\'e}ndez-Delmestre}, {Mizusawa}, {Mu{\~n}oz-Mateos}, {Regan}, {Salo}, {Seibert}, \& {Sheth}}]{2012MNRAS.427.1102M}
{Mart{\'\i}n-Navarro}, I., {Bakos}, J., {Trujillo}, I., {et~al.} 2012{\natexlab{b}}, \mnras, 427, 1102

\bibitem[{{Mart{\'\i}nez-Lombilla} {et~al.}(2019){Mart{\'\i}nez-Lombilla}, {Trujillo}, \& {Knapen}}]{martinezlombilla2019}
{Mart{\'\i}nez-Lombilla}, C., {Trujillo}, I., \& {Knapen}, J.~H. 2019, \mnras, 483, 664

\bibitem[{{Mart{\'\i}nez-Serrano} {et~al.}(2009){Mart{\'\i}nez-Serrano}, {Serna}, {Dom{\'e}nech-Moral}, \& {Dom{\'\i}nguez-Tenreiro}}]{martinez2009}
{Mart{\'\i}nez-Serrano}, F.~J., {Serna}, A., {Dom{\'e}nech-Moral}, M., \& {Dom{\'\i}nguez-Tenreiro}, R. 2009, \apjl, 705, L133

\bibitem[{{Mendel} {et~al.}(2014){Mendel}, {Simard}, {Palmer}, {Ellison}, \& {Patton}}]{2014ApJS..210....3M}
{Mendel}, J.~T., {Simard}, L., {Palmer}, M., {Ellison}, S.~L., \& {Patton}, D.~R. 2014, \apjs, 210, 3

\bibitem[{{Merritt} {et~al.}(2016){Merritt}, {van Dokkum}, {Abraham}, \& {Zhang}}]{2016ApJ...830...62M}
{Merritt}, A., {van Dokkum}, P., {Abraham}, R., \& {Zhang}, J. 2016, \apj, 830, 62

\bibitem[{{Mihos} {et~al.}(2013){Mihos}, {Harding}, {Spengler}, {Rudick}, \& {Feldmeier}}]{2013ApJ...762...82M}
{Mihos}, J.~C., {Harding}, P., {Spengler}, C.~E., {Rudick}, C.~S., \& {Feldmeier}, J.~J. 2013, \apj, 762, 82

\bibitem[{{Nair} \& {Abraham}(2010)}]{nair}
{Nair}, P.~B. \& {Abraham}, R.~G. 2010, \apjs, 186, 427

\bibitem[{Oke \& Gunn(1983)}]{oke1983}
Oke, J. \& Gunn, J. 1983, Astrophysical Journal, Part 1, vol. 266, Mar. 15, 1983, p. 713-717., 266, 713

\bibitem[{Pohlen \& Trujillo(2006)}]{pohlen2006}
Pohlen, M. \& Trujillo, I. 2006, Astronomy \& Astrophysics, 454, 759

\bibitem[{{Pohlen} \& {Trujillo}(2006)}]{Pohlen_2006}
{Pohlen}, M. \& {Trujillo}, I. 2006, \aap, 454, 759

\bibitem[{{Rich} {et~al.}(2019){Rich}, {Mosenkov}, {Lee-Saunders}, {Koch}, {Kormendy}, {Kennefick}, {Brosch}, {Sales}, {Bullock}, {Burkert}, {Collins}, {Cooper}, {Fusco}, {Reitzel}, {Thilker}, {Milewski}, {Elias}, {Saade}, \& {De Groot}}]{2019MNRAS.490.1539R}
{Rich}, R.~M., {Mosenkov}, A., {Lee-Saunders}, H., {et~al.} 2019, \mnras, 490, 1539

\bibitem[{{Roediger} \& {Courteau}(2015)}]{2015MNRAS.452.3209R}
{Roediger}, J.~C. \& {Courteau}, S. 2015, \mnras, 452, 3209

\bibitem[{{Rom{\'a}n} \& {Trujillo}(2018)}]{2018RNAAS...2..144R}
{Rom{\'a}n}, J. \& {Trujillo}, I. 2018, Research Notes of the American Astronomical Society, 2, 144

\bibitem[{{Ro{\v{s}}kar} {et~al.}(2008){Ro{\v{s}}kar}, {Debattista}, {Stinson}, {Quinn}, {Kaufmann}, \& {Wadsley}}]{2008ApJ...675L..65R}
{Ro{\v{s}}kar}, R., {Debattista}, V.~P., {Stinson}, G.~S., {et~al.} 2008, \apjl, 675, L65

\bibitem[{{Ruiz-Lara} {et~al.}(2016){Ruiz-Lara}, {P{\'e}rez}, {Florido}, {S{\'a}nchez-Bl{\'a}zquez}, {M{\'e}ndez-Abreu}, {Lyubenova}, {Falc{\'o}n-Barroso}, {S{\'a}nchez-Menguiano}, {S{\'a}nchez}, {Galbany}, {Garc{\'\i}a-Benito}, {Gonz{\'a}lez Delgado}, {Husemann}, {Kehrig}, {L{\'o}pez-S{\'a}nchez}, {Marino}, {Mast}, {Papaderos}, {van de Ven}, {Walcher}, {Zibetti}, \& {CALIFA Team}}]{Ruiz-Lara2016}
{Ruiz-Lara}, T., {P{\'e}rez}, I., {Florido}, E., {et~al.} 2016, \mnras, 456, L35

\bibitem[{{S{\'a}nchez-Bl{\'a}zquez} {et~al.}(2009){S{\'a}nchez-Bl{\'a}zquez}, {Courty}, {Gibson}, \& {Brook}}]{sanchez2009}
{S{\'a}nchez-Bl{\'a}zquez}, P., {Courty}, S., {Gibson}, B.~K., \& {Brook}, C.~B. 2009, \mnras, 398, 591

\bibitem[{{Teeninga} {et~al.}(2016){Teeninga}, {Moschini}, {Trager}, \& {Wilkinson}}]{Teeninga2016}
{Teeninga}, P., {Moschini}, U., {Trager}, S.~C., \& {Wilkinson}, M.~H.~F. 2016, Math. Morphol. Theory Appl., 1, 25

\bibitem[{{Trujillo} {et~al.}(2020){Trujillo}, {Chamba}, \& {Knapen}}]{Trujillo2020}
{Trujillo}, I., {Chamba}, N., \& {Knapen}, J.~H. 2020, \mnras, 493, 87

\bibitem[{{Trujillo} {et~al.}(2021){Trujillo}, {D'Onofrio}, {Zaritsky}, {Madrigal-Aguado}, {Chamba}, {Golini}, {Akhlaghi}, {Sharbaf}, {Infante-Sainz}, {Rom{\'a}n}, {Morales-Socorro}, {Sand}, \& {Martin}}]{LIGHTSs}
{Trujillo}, I., {D'Onofrio}, M., {Zaritsky}, D., {et~al.} 2021, \aap, 654, A40

\bibitem[{Trujillo {et~al.}(2021)Trujillo, D’Onofrio, Zaritsky, Madrigal-Aguado, Chamba, Golini, Akhlaghi, Sharbaf, Infante-Sainz, Rom{\'a}n, {et~al.}}]{trujillo2021}
Trujillo, I., D’Onofrio, M., Zaritsky, D., {et~al.} 2021, Astronomy \& Astrophysics, 654, A40

\bibitem[{{Trujillo} \& {Fliri}(2016)}]{Trujillo2016}
{Trujillo}, I. \& {Fliri}, J. 2016, \apj, 823, 123

\bibitem[{{van der Kruit}(1979)}]{1979A&AS...38...15V}
{van der Kruit}, P.~C. 1979, \aaps, 38, 15

\bibitem[{Watkins {et~al.}(2016)Watkins, Mihos, \& Harding}]{Watkins_2016}
Watkins, A.~E., Mihos, J.~C., \& Harding, P. 2016, The Astrophysical Journal, 826, 59

\bibitem[{{York} {et~al.}(2000){York}, {Adelman}, {Anderson}, {Anderson}, {Annis}, {Bahcall}, {Bakken}, {Barkhouser}, {Bastian}, {Berman}, {Boroski}, {Bracker}, {Briegel}, {Briggs}, {Brinkmann}, {Brunner}, {Burles}, {Carey}, {Carr}, {Castander}, {Chen}, {Colestock}, {Connolly}, {Crocker}, {Csabai}, {Czarapata}, {Davis}, {Doi}, {Dombeck}, {Eisenstein}, {Ellman}, {Elms}, {Evans}, {Fan}, {Federwitz}, {Fiscelli}, {Friedman}, {Frieman}, {Fukugita}, {Gillespie}, {Gunn}, {Gurbani}, {de Haas}, {Haldeman}, {Harris}, {Hayes}, {Heckman}, {Hennessy}, {Hindsley}, {Holm}, {Holmgren}, {Huang}, {Hull}, {Husby}, {Ichikawa}, {Ichikawa}, {Ivezi{\'c}}, {Kent}, {Kim}, {Kinney}, {Klaene}, {Kleinman}, {Kleinman}, {Knapp}, {Korienek}, {Kron}, {Kunszt}, {Lamb}, {Lee}, {Leger}, {Limmongkol}, {Lindenmeyer}, {Long}, {Loomis}, {Loveday}, {Lucinio}, {Lupton}, {MacKinnon}, {Mannery}, {Mantsch}, {Margon}, {McGehee}, {McKay}, {Meiksin}, {Merelli}, {Monet}, {Munn}, {Narayanan}, {Nash}, {Neilsen}, {Neswold}, {Newberg}, {Nichol}, {Nicinski},
  {Nonino}, {Okada}, {Okamura}, {Ostriker}, {Owen}, {Pauls}, {Peoples}, {Peterson}, {Petravick}, {Pier}, {Pope}, {Pordes}, {Prosapio}, {Rechenmacher}, {Quinn}, {Richards}, {Richmond}, {Rivetta}, {Rockosi}, {Ruthmansdorfer}, {Sandford}, {Schlegel}, {Schneider}, {Sekiguchi}, {Sergey}, {Shimasaku}, {Siegmund}, {Smee}, {Smith}, {Snedden}, {Stone}, {Stoughton}, {Strauss}, {Stubbs}, {SubbaRao}, {Szalay}, {Szapudi}, {Szokoly}, {Thakar}, {Tremonti}, {Tucker}, {Uomoto}, {Vanden Berk}, {Vogeley}, {Waddell}, {Wang}, {Watanabe}, {Weinberg}, {Yanny}, {Yasuda}, \& {SDSS Collaboration}}]{2000AJ....120.1579Y}
{York}, D.~G., {Adelman}, J., {Anderson}, John~E., J., {et~al.} 2000, \aj, 120, 1579

\bibitem[{{Zaritsky} {et~al.}(2024){Zaritsky}, {Golini}, {Donnerstein}, {Trujillo}, {Akhlaghi}, {Chamba}, {D'Onofrio}, {Eskandarlou}, {Hosseini-ShahiSavandi}, {Infante-Sainz}, {Martin}, {Montes}, {Rom{\'a}n}, {Sedighi}, \& {Sharbaf}}]{2024Zaritskydennislights}
{Zaritsky}, D., {Golini}, G., {Donnerstein}, R., {et~al.} 2024, \aj, 168, 69

\bibitem[{{Zheng} {et~al.}(2015){Zheng}, {Thilker}, {Heckman}, {Meurer}, {Burgett}, {Chambers}, {Huber}, {Kaiser}, {Magnier}, {Metcalfe}, {Price}, {Tonry}, {Wainscoat}, \& {Waters}}]{Zheng2015}
{Zheng}, Z., {Thilker}, D.~A., {Heckman}, T.~M., {et~al.} 2015, \apj, 800, 120

\end{thebibliography}

\clearpage
\onecolumn
\appendix
\section{The profiles of selected galaxies }
\label{profiles}

In this appendix, we show the color images of the galaxies in the sample together with the surface brightness profiles of the objects in the $g$ and $r$ Sloan bands. The color images were created using the \texttt{astscript-radial-profile} \citep{astscript-color-faint-gray}, which is part of the Gnuastro program \cite{gnuastro}, using the $g$, $r$, and $i$ Sloan filters. The code uses the average of the 3 filters for the gray background pixels. The stellar mass profiles are also shown along with the color profiles. All these profiles (which include Galactic extinction, galaxy inclination, and cosmological dimming correction) are taken directly from the work of \citet{Chamba_2022}.

\begin{figure*}[h]
    \centering
    \includegraphics[width=0.69\linewidth, height=0.7\linewidth]{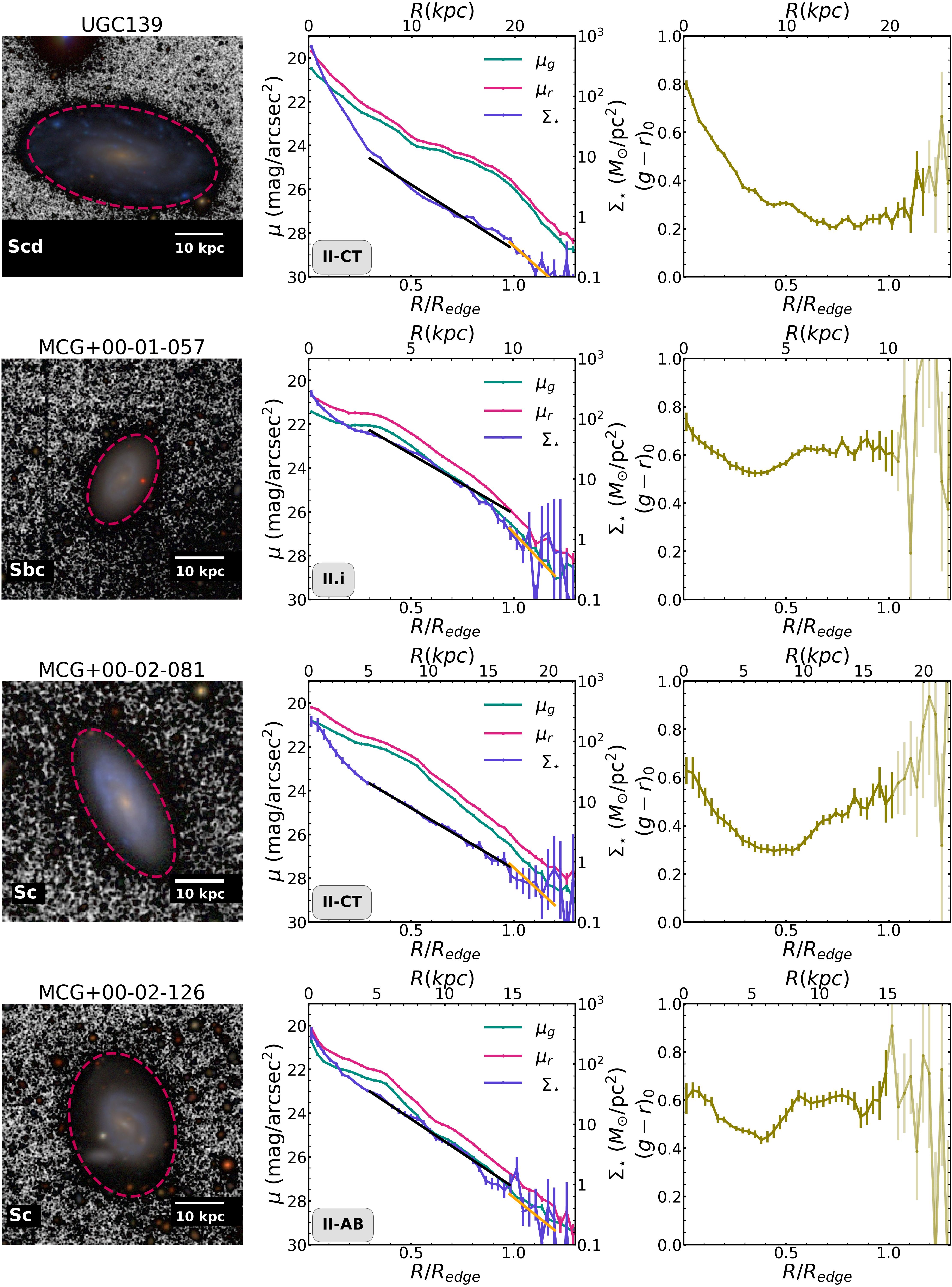}
    \caption{Galaxy color image and radial profiles. The left column shows a composite color image of the galaxy using the $g$, $r$, and $i$ Sloan bands. The background is shown in black and white to visualize the low surface brightness features better. The name of the object, its morphological type, and a 10 kpc length bar for reference are superimposed. The middle column displays the surface brightness profiles of the galaxies in the $g$ and $r$ bands. These profiles are obtained using elliptical apertures with an axis ratio and position angle as indicated by the red dashed ellipse in the galaxy image (left column). The semi-major axis of the red dashed ellipse corresponds to R$_{edge}$\citep{Chamba_2022}. The surface brightness profiles are corrected for Galactic extinction, galaxy inclination, and cosmological dimming.  The surface stellar mass profile is also added. The exponential fittings to the disk inside the edge (black line) and to the region beyond the edge of the galaxy (orange line) are included. The right column shows the Galactic extinction-corrected color profile of the galaxy.}
    \label{galaxysample}
\end{figure*}

\begin{figure*}
    \centering
    \setcounter{figure}{0}
    \includegraphics[width=0.64\linewidth, height=0.65\linewidth]{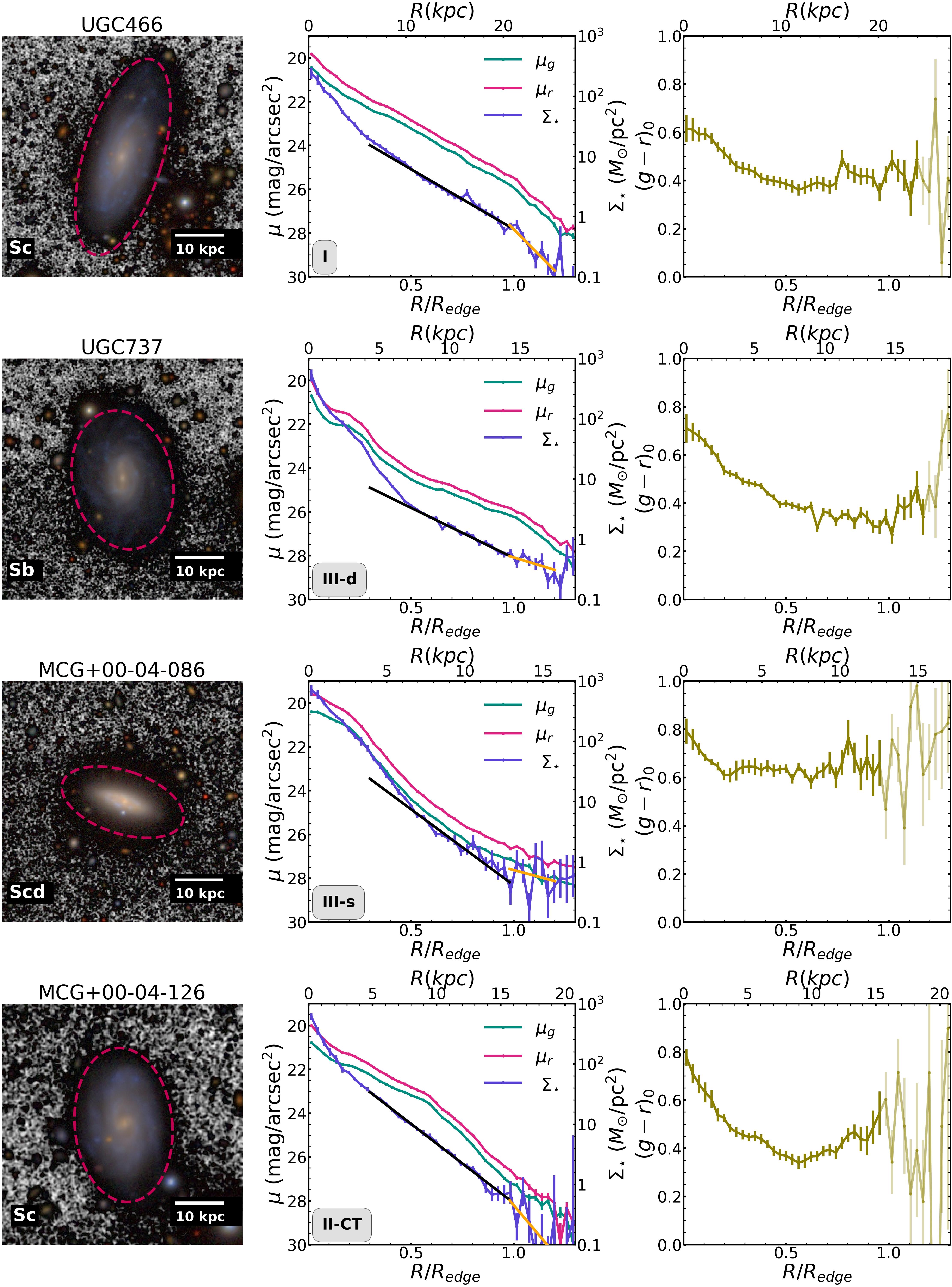}
        \includegraphics[width=0.64\linewidth, height=0.65\linewidth]{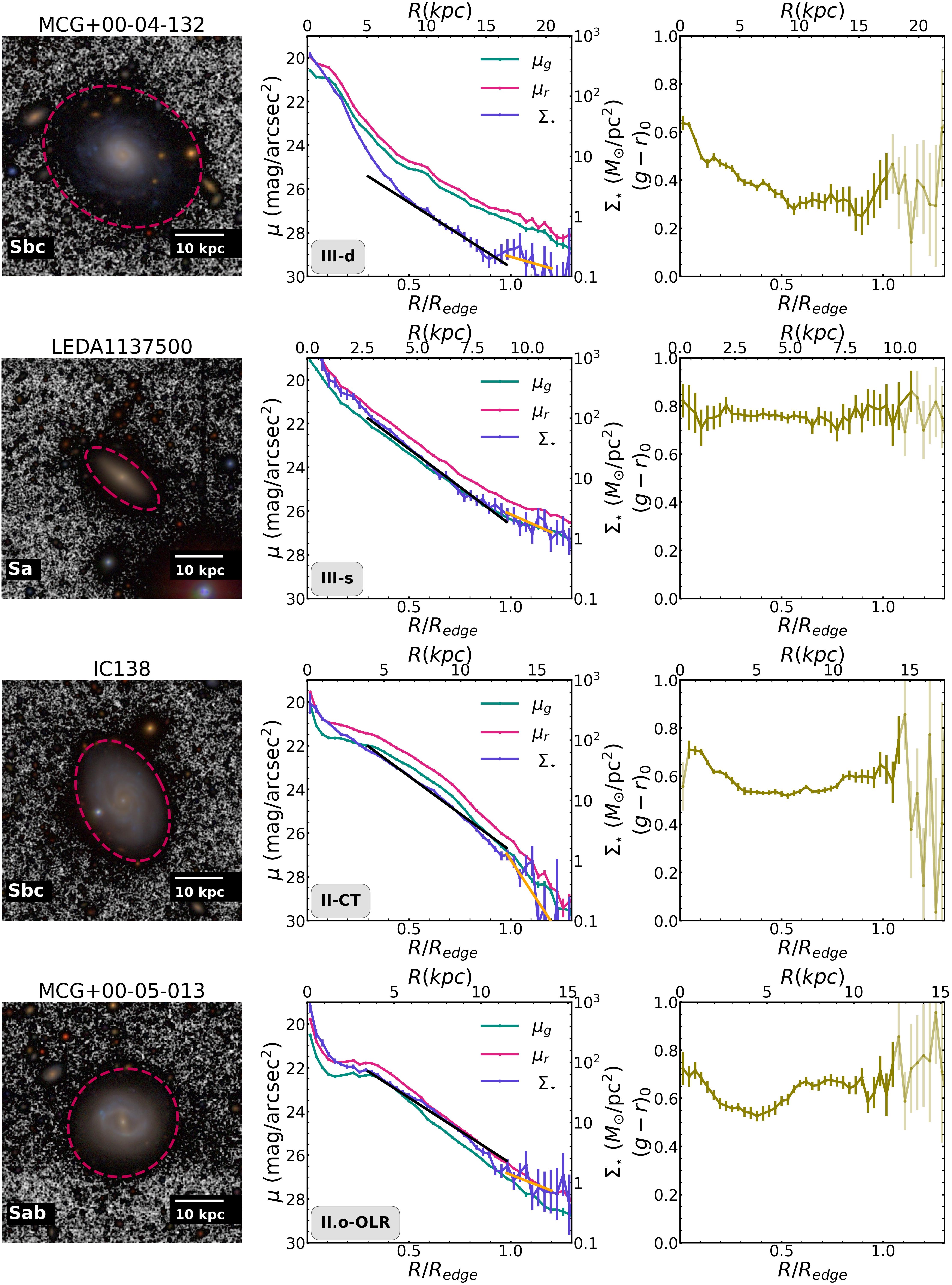}
    \caption{continued.}   
    \label{}
\end{figure*}

\begin{figure*}
    \centering
    \setcounter{figure}{0}
    \includegraphics[width=0.64\linewidth, height=0.65\linewidth]{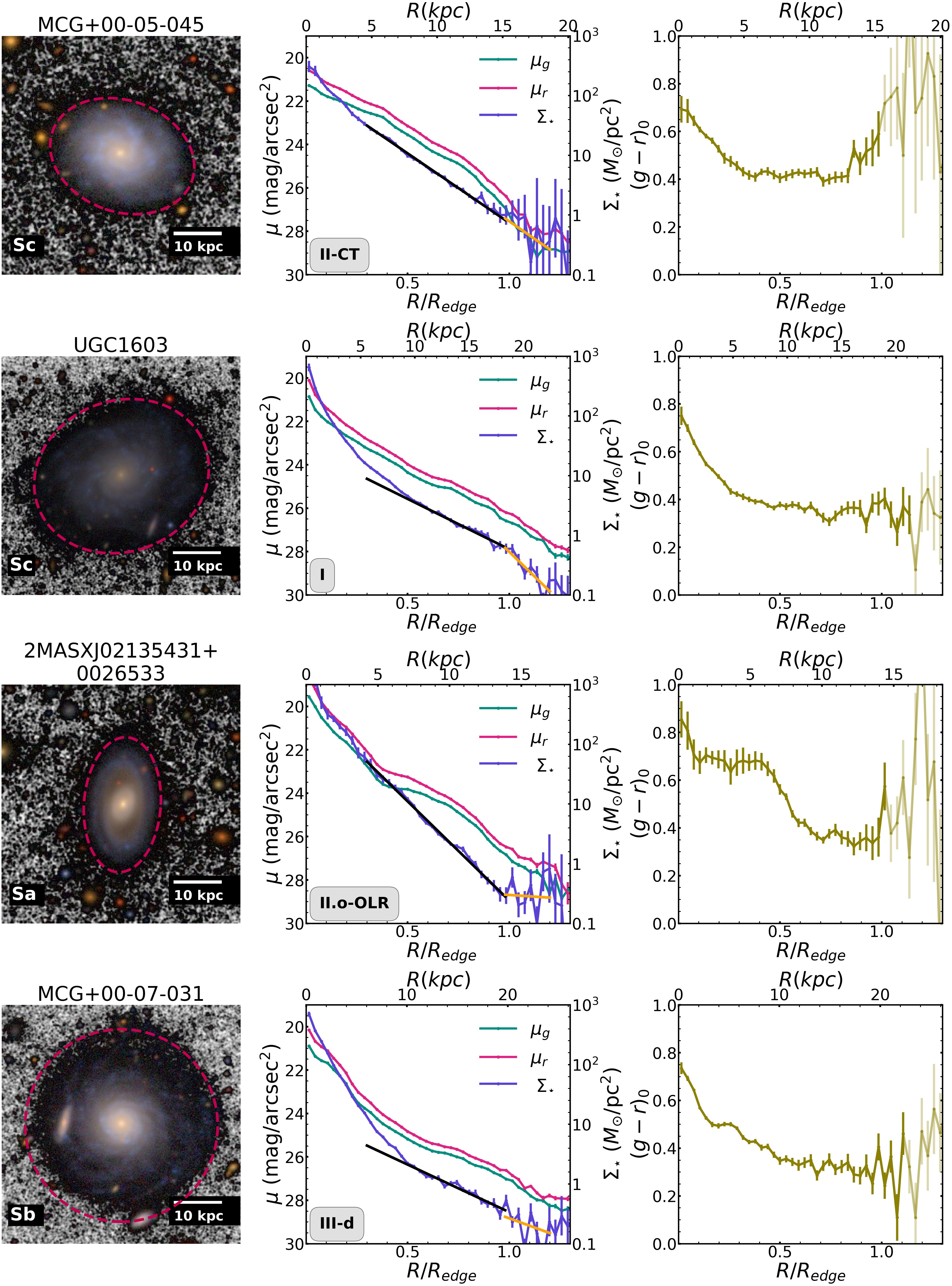}
        \includegraphics[width=0.64\linewidth, height=0.65\linewidth]{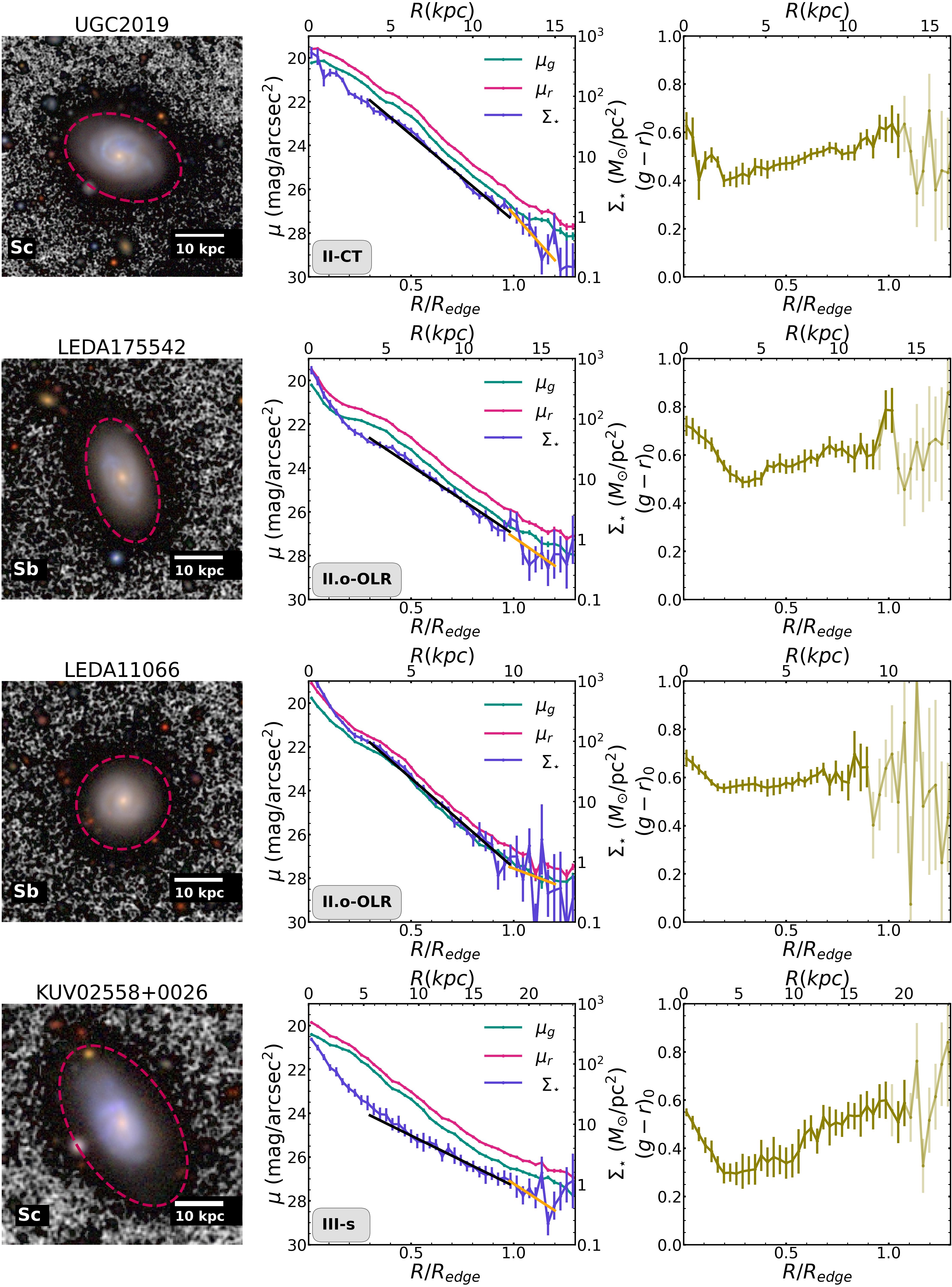}
    \caption{continued.}
    \label{}
\end{figure*}

\begin{figure*}
    \centering
    \setcounter{figure}{0}
    \includegraphics[width=0.64\linewidth, height=0.65\linewidth]{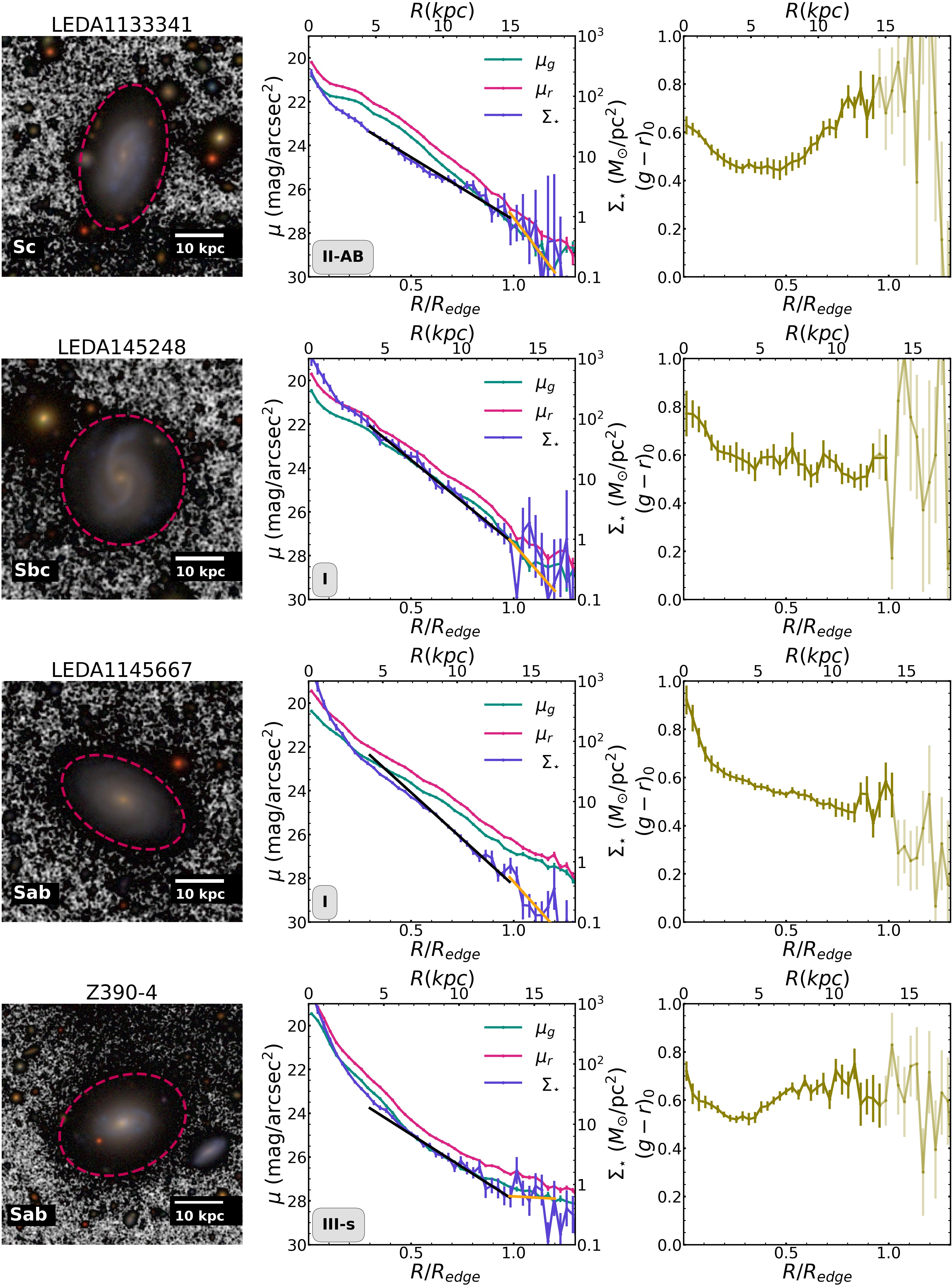}
        \includegraphics[width=0.64\linewidth, height=0.65\linewidth]{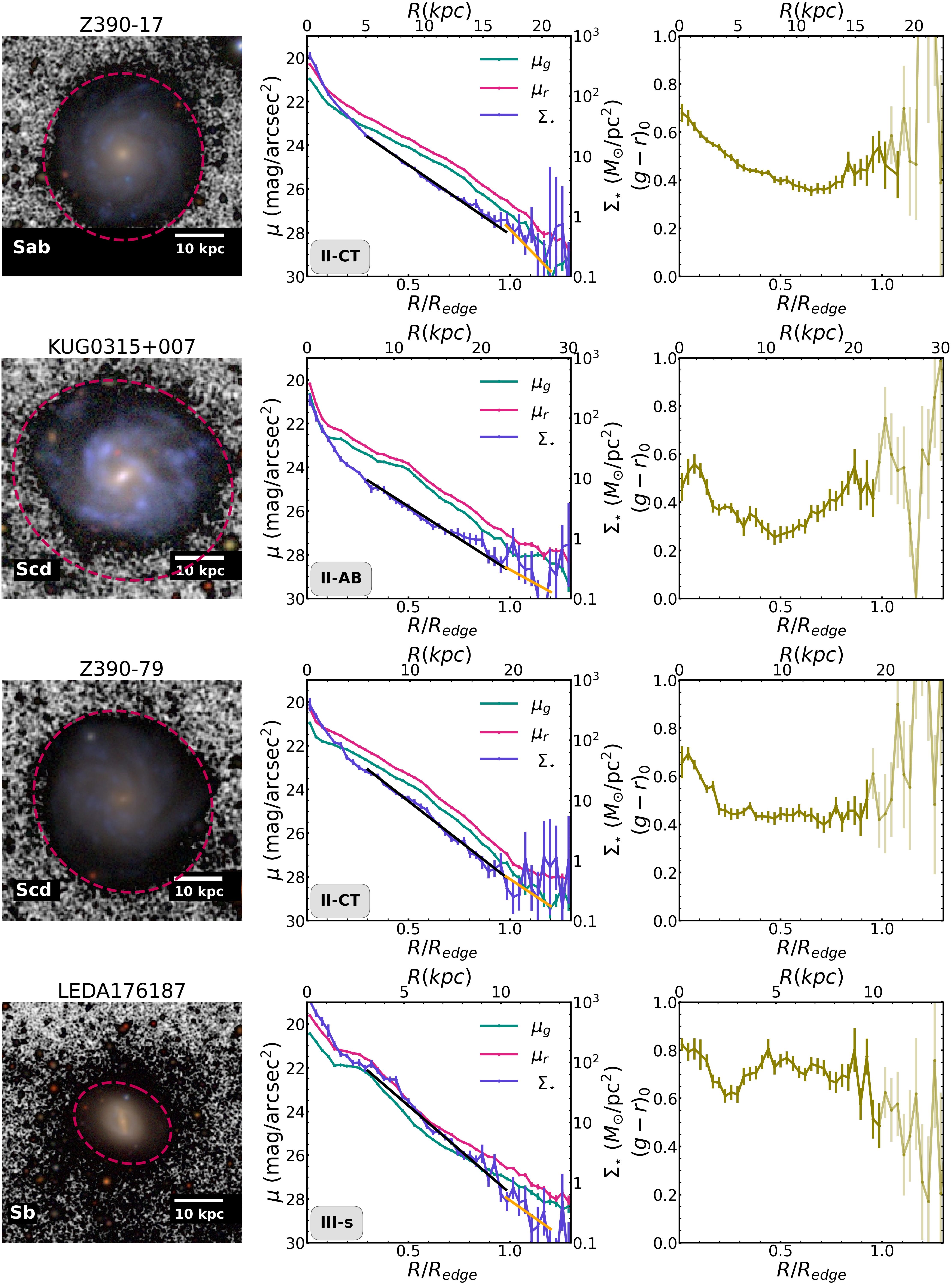}
    \caption{continued.}
    \label{}
\end{figure*}

\begin{figure*}
    \centering
    \setcounter{figure}{0}
    \includegraphics[width=0.64\linewidth, height=0.65\linewidth]{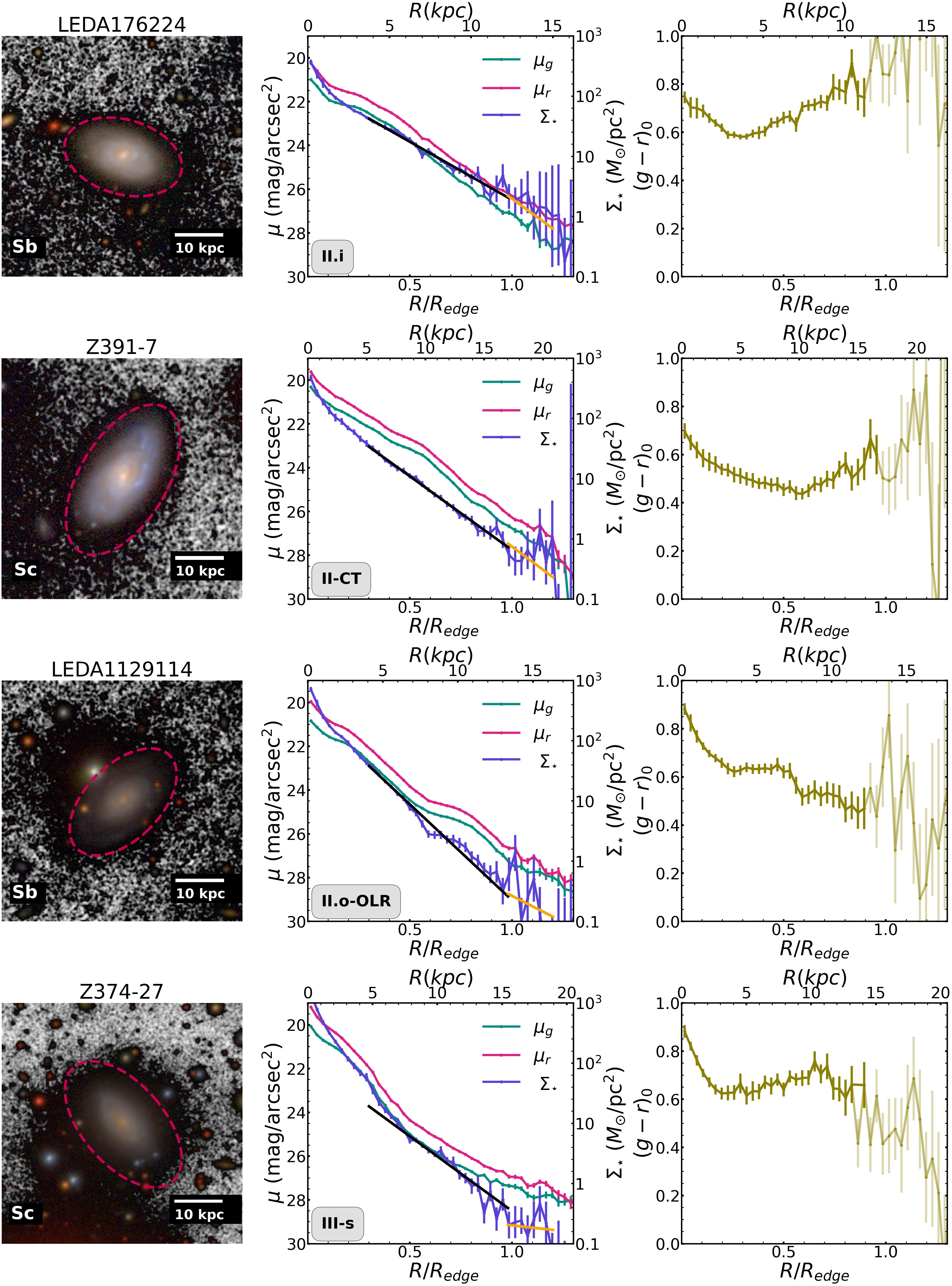}
        \includegraphics[width=0.64\linewidth, height=0.65\linewidth]{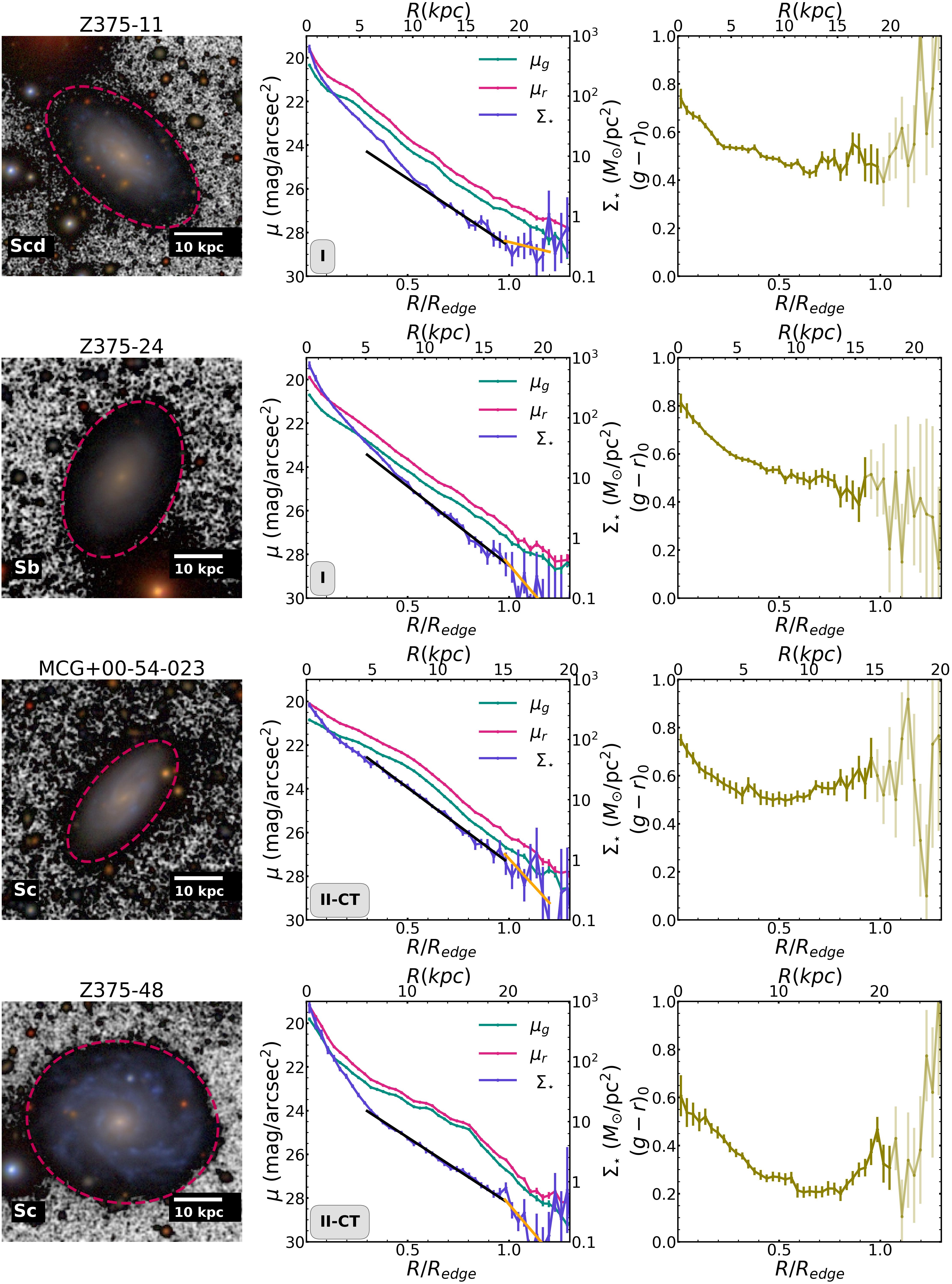}
    \caption{continued.}
    \label{}
\end{figure*}

\begin{figure*}
    \centering
    \setcounter{figure}{0}
    \includegraphics[width=0.64\linewidth, height=0.65\linewidth]{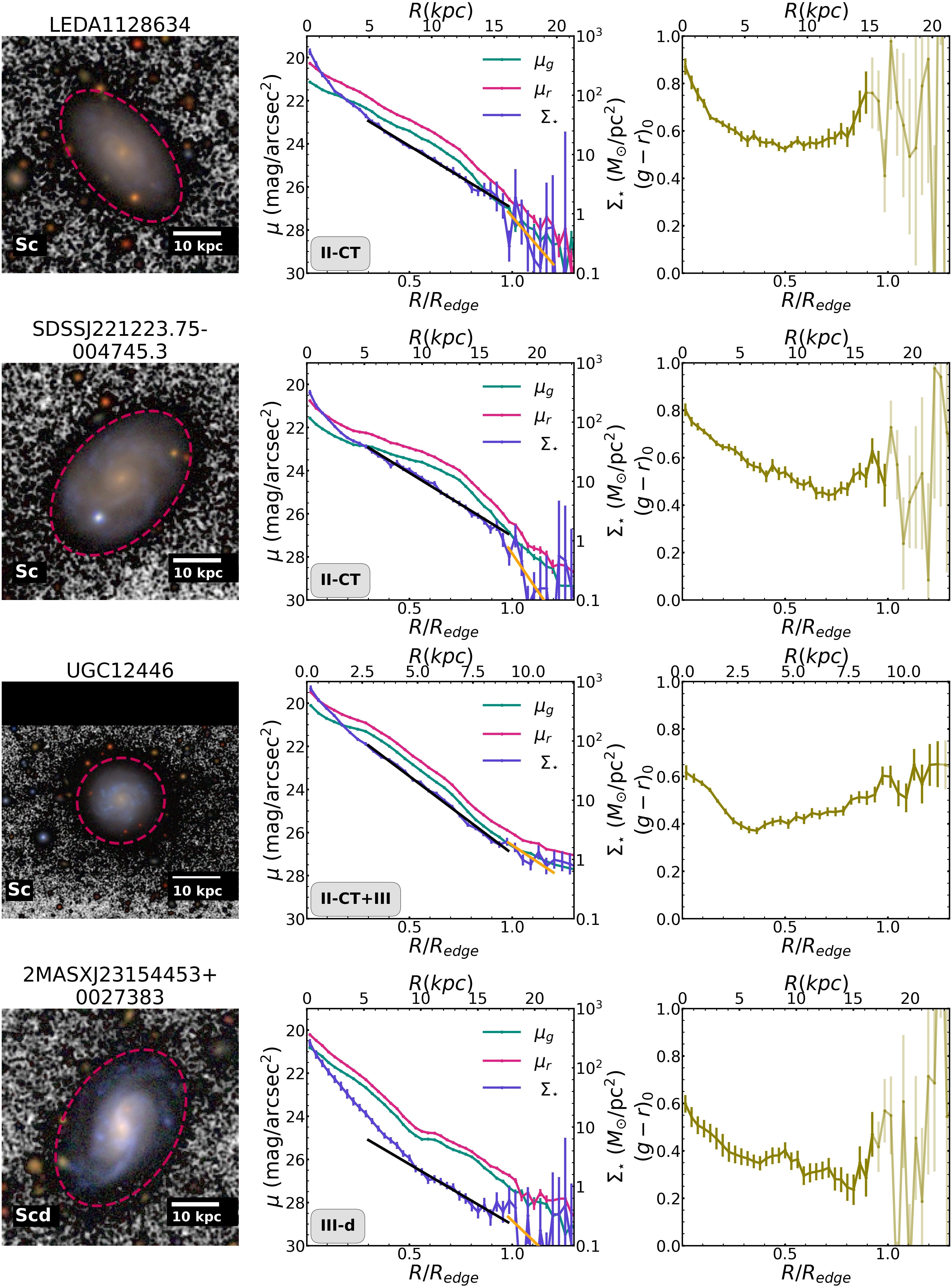}
    \includegraphics[width=0.64\linewidth, height=0.65\linewidth]{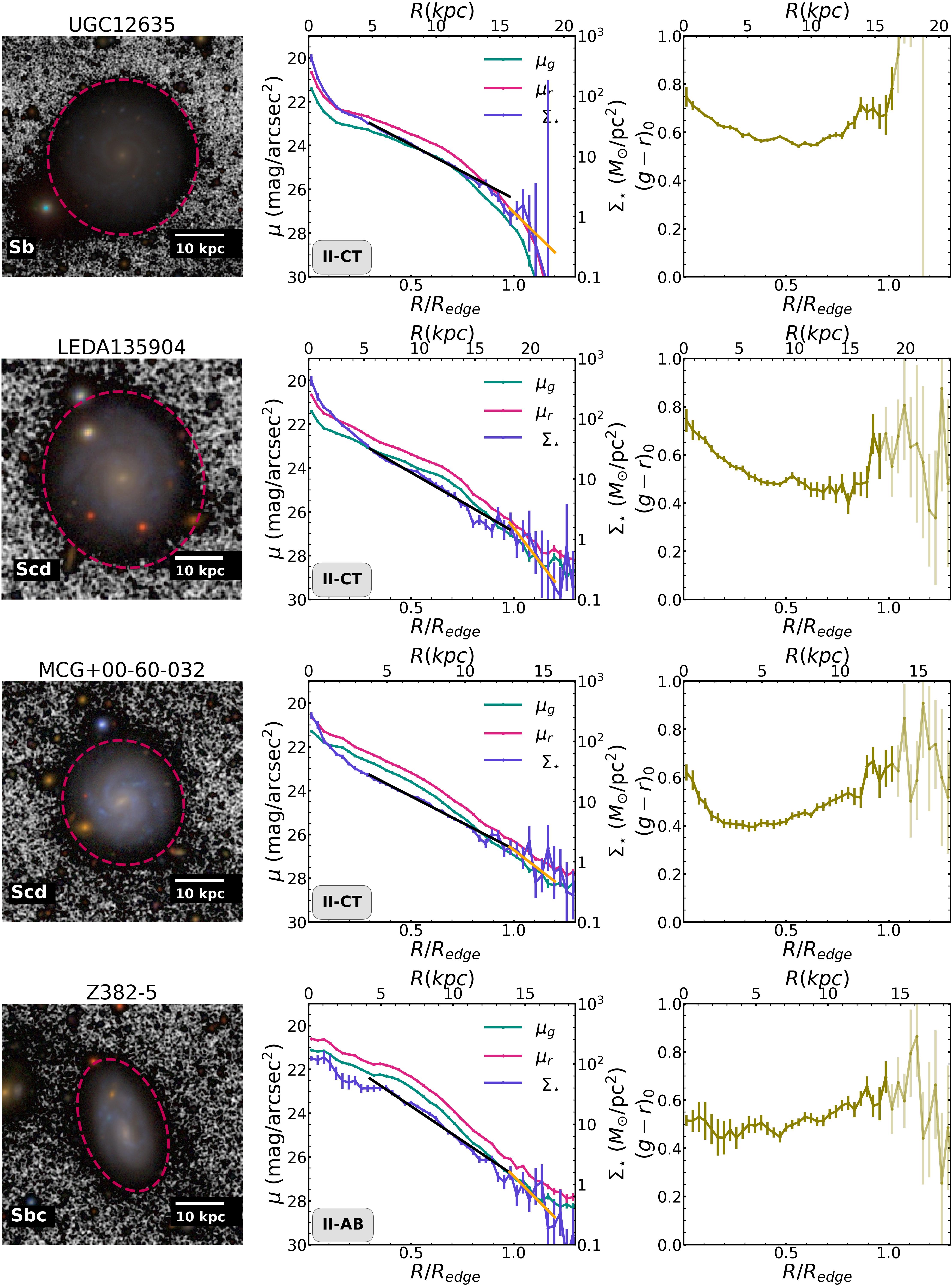}
    \caption{continued.}
    \label{}
\end{figure*}

\newpage
\begin{table*}[]
\caption{Characteristics of the galaxies studied in this work. }
\label{galaxyproperty}
\begin{tabular}{|l|S[table-format=3.5]|S[table-format=+2.5]|l|S[table-format=1.4]|l|S[table-format=2.2]|S[table-format=1.2]|}
\hline
\hline
Name & \multicolumn{1}{c|}{RA (J2000)} & \multicolumn{1}{c|}{Dec (J2000)} & Morphology & \multicolumn{1}{c|}{Redshift} & Type & \multicolumn{1}{c|}{Total Mass } & b/a \\ 
 & \multicolumn{1}{c|}{(deg)} & \multicolumn{1}{c|}{(deg)} &  & \multicolumn{1}{c|}{} &  &  \multicolumn{1}{c|}{($\times10^{10}$ M$_{\odot}$)} &  \\ \hline

UGC139                  & 3.63278   & -0.73761   & Scd & 0.0140 & II-CT      & 0.95 & 0.51 \\ \hline
MCG+00-01-057           & 4.28682   & -0.95825   & Sbc & 0.0189 & II.i       & 0.66 & 0.62 \\ \hline
MCG+00-02-081           & 8.78594   & -0.13351   & Sc  & 0.0302 & II-CT      & 0.80 & 0.44 \\ \hline
MCG+00-02-126           & 10.44259  & 1.17708    & Sc  & 0.0180 & II-AB      & 0.92 & 0.71 \\ \hline
UGC466                  & 10.96603  & 0.80208    & Sc  & 0.0181 & I          & 1.16 & 0.39 \\ \hline
UGC737                  & 17.71186  & 0.19805    & Sb  & 0.0176 & III-d      & 0.64 & 0.71 \\ \hline
MCG+00-04-086           & 20.03265  & 0.07698    & Scd & 0.0186 & III-s      & 1.24 & 0.50 \\ \hline
MCG+00-04-126           & 21.23027  & 0.21150    & Sc  & 0.0285 & II-CT      & 1.07 & 0.63 \\ \hline
MCG+00-04-132           & 21.41562  & 1.17792    & Sbc & 0.0195 & III-d      & 0.76 & 0.81 \\ \hline
LEDA1137500             & 22.84564  & -0.67440   & Sa  & 0.0157 & III-s      & 1.25 & 0.42 \\ \hline
IC138                   & 23.25846  & -0.68979   & Sbc & 0.0154 & II-CT      & 1.30 & 0.66 \\ \hline
MCG+00-05-013           & 23.90285  & 0.01481    & Sab & 0.0173 & II.o-OLR   & 1.01 & 0.93 \\ \hline
MCG+00-05-045           & 28.52577  & 0.95036    & Sc  & 0.0275 & II-CT      & 0.96 & 0.75 \\ \hline
UGC1603                 & 31.67721  & -0.86055   & Sc  & 0.0200 & I          & 0.87 & 0.85 \\ \hline
2MASXJ02135431+0026533  & 33.47645  & 0.44827    & Sa  & 0.0248 & II.o-OLR   & 1.56 & 0.57 \\ \hline
MCG+00-07-031           & 37.93003  & 0.90435    & Sb  & 0.0210 & III-d      & 1.04 & 1.00 \\ \hline
UGC2019                 & 38.16328  & 0.61730    & Sc  & 0.0207 & II-CT      & 1.21 & 0.68 \\ \hline
LEDA175542              & 42.57503  & -0.35576   & Sb  & 0.0289 & II.o-OLR   & 1.03 & 0.53 \\ \hline
LEDA11066               & 43.94556  & -0.10428   & Sb  & 0.0292 & II.o-OLR   & 1.04 & 0.95 \\ \hline
KUV02558+0026           & 44.60194  & 0.63750    & Sc  & 0.0403 & III-s      & 1.07 & 0.56 \\ \hline
LEDA1133341             & 46.16470  & -0.84142   & Sc  & 0.0275 & II-AB      & 0.71 & 0.57 \\ \hline
LEDA145248              & 46.16994  & -0.12024   & Sbc & 0.0311 & I          & 1.55 & 0.95 \\ \hline
LEDA1145667             & 46.33151  & -0.34506   & Sab & 0.0293 & I          & 1.12 & 0.59 \\ \hline
Z390-4                  & 46.95629  & -0.78365   & Sab & 0.0197 & III-s      & 0.92 & 0.72 \\ \hline
Z390-17                 & 47.88913  & -0.74315   & Sab & 0.0282 & II-CT      & 0.90 & 0.95 \\ \hline
KUG0315+007             & 49.64145  & 0.92536    & Scd & 0.0359 & II-AB      & 0.72 & 0.86 \\ \hline
Z390-79                 & 50.84990  & 0.44532    & Scd & 0.0353 & II-CT      & 1.42 & 0.91 \\ \hline
LEDA176187              & 51.69435  & -0.64829   & Sb  & 0.0205 & III-s      & 0.95 & 0.76 \\ \hline
LEDA176224              & 51.87848  & -1.12786   & Sb  & 0.0259 & II.i       & 0.86 & 0.62 \\ \hline
Z391-7                  & 53.18565  & -1.06886   & Sc  & 0.0308 & II-CT      & 1.25 & 0.52 \\ \hline
LEDA1129114             & 53.46165  & -1.00847   & Sb  & 0.0260 & II.o-OLR   & 0.84 & 0.57 \\ \hline
Z374-27                 & 313.83036 & 0.54049    & Sc  & 0.0232 & III-s      & 1.50 & 0.54 \\ \hline
Z375-11                 & 318.30435 & -0.36667   & Scd & 0.0204 & I          & 1.27 & 0.54 \\ \hline
Z375-24                 & 319.09161 & 1.08483    & Sb  & 0.0316 & I          & 1.36 & 0.66 \\ \hline
MCG+00-54-023           & 322.26290 & -0.47368   & Sc  & 0.0300 & II-CT      & 1.22 & 0.47 \\ \hline
Z375-48                 & 322.74928 & -0.00060   & Sc  & 0.0301 & II-CT      & 1.33 & 0.84 \\ \hline
LEDA1128634             & 332.80426 & -1.02716   & Sc  & 0.0338 & II-CT      & 1.38 & 0.55 \\ \hline
SDSSJ221223.75-004745.3 & 333.09911 & -0.79613   & Sc  & 0.0338 & II-CT      & 1.32 & 0.64 \\ \hline
UGC12446                & 348.63821 & 0.23582    & Sc  & 0.0147 & II-CT+III  & 0.84 & 0.97 \\ \hline
2MASXJ23154453+0027383  & 348.93574 & 0.46046    & Scd & 0.0345 & III-d      & 0.65 & 0.65 \\ \hline
UGC12635                & 352.60699 & 0.15653    & Sb  & 0.0174 & II-CT      & 1.10 & 0.97 \\ \hline
LEDA135904              & 354.54601 & -0.46309   & Scd & 0.0352 & II-CT      & 1.49 & 0.90 \\ \hline
MCG+00-60-032           & 355.13611 & -0.55103   & Scd & 0.0231 & II-CT      & 0.64 & 0.94 \\ \hline
Z382-5                  & 359.05660 & -0.53947   & Sbc & 0.0222 & II-AB      & 1.00 & 0.60 \\ \hline
\end{tabular}
\tablefoot{The RA and DEC coordinates, redshift values, total stellar mass, and axis ratio (b/a) of the outer disk are extracted from the catalog presented in \citet{Chamba_2022}. The morphological classification is obtained from \citet{nair}. The profile Type is estimated in this work using as a criterion the shape of the surface brightness profiles down to the limiting surface brightness of the SDSS (i.e. $\mu_g$$\sim$26.5 mag/arcsec$^2$).}
\end{table*}

\end{document}